\documentclass[aps,prx,superscriptaddress,twocolumn,notitlepage,bibnotes, longbibliography]{revtex4-1}

\usepackage{amsthm,amsmath,amssymb,amsbsy}

\usepackage{url}

\usepackage{graphicx}
\usepackage{bm,bbm}%
\usepackage{physics}
\usepackage{enumerate}
\usepackage{booktabs,multirow}
\usepackage{mathtools,bbding}
\usepackage{microtype}
\usepackage{mathrsfs}

\usepackage{txfonts,newtxmath}

\usepackage[most,breakable]{tcolorbox}

\newcommand{\nc}{\newcommand}
\newcommand{\rnc}{\renewcommand}

\makeatletter
\newcommand*\rel@kern[1]{\kern#1\dimexpr\macc@kerna}
\newcommand*\widebar[1]{%
	\begingroup
	\def\mathaccent##1##2{%
		\rel@kern{0.8}%
		\overline{\rel@kern{-0.8}\macc@nucleus\rel@kern{0.2}}%
		\rel@kern{-0.2}%
	}%
	\macc@depth\@ne
	\let\math@bgroup\@empty \let\math@egroup\macc@set@skewchar
	\mathsurround\z@ \frozen@everymath{\mathgroup\macc@group\relax}%
	\macc@set@skewchar\relax
	\let\mathaccentV\macc@nested@a
	\macc@nested@a\relax111{#1}%
	\endgroup
}

\rnc{\thesection}{\arabic{section}}
\rnc{\thesubsection}{\thesection.\arabic{subsection}}
\rnc{\thesubsubsection}{\thesubsection.\arabic{subsubsection}}

\usepackage{etoolbox}
\usepackage{hyperref}
\hypersetup{linktoc=all}

\newtheorem{manualtheoreminner}{Theorem}
\newenvironment{manualtheorem}[1]{%
  \renewcommand\themanualtheoreminner{#1}%
  \manualtheoreminner
}{\endmanualtheoreminner}

\DeclareMathOperator{\supp}{supp}

\nc{\tcr}[1]{{\color{red} #1}}
\nc{\tcb}[1]{{\color{blue} #1}}

\nc{\psic}{\psi^{c}}
\nc{\two}[1]{\underline{2^{d-#1}}}
\rnc{\H}{\mathcal{H}}
\nc{\Hanc}{\mathcal{H}_{\text{anc}}}
\nc{\psianc}{\psi_{\text{anc}}}
\nc{\lampow}{\lambda^{1/d}}

\nc{\proj}[1]{\ket{#1}\!\bra{#1}}
\nc{\pro}[1]{#1 #1^\dagger}
\nc{\RR}{{{\mathbb R}}}
\nc{\CC}{{{\mathbb C}}}
\nc{\FF}{{{\mathbb F}}}
\nc{\NN}{{{\mathbb N}}}
\nc{\ZZ}{{{\mathbb Z}}}

\nc{\MIO}{{\text{\rm MIO}}}
\nc{\DIO}{{\text{\rm DIO}}}
\nc{\SIO}{{\text{\rm SIO}}}
\nc{\IO}{{\text{\rm IO}}}

\nc{\SEP}{{\text{SEP}}}
\nc{\NS}{{\text{NS}}}
\nc{\LOCC}{{\text{LOCC}}}
\nc{\PPT}{{\text{PPT}}}
\nc{\EXT}{{\text{EXT}}}
\nc{\OLOCC}{{\text{1-LOCC}}}
\nc{\SEPP}{{\text{SEPP}}}

\nc{\MC}{{\text{\rm MC}}}

\nc{\cE}{\mathscr{E}}

\rnc{\*}{\textup{*}}
\newcommand{\mc}[1]{\mathcal{#1}}
\newcommand{\bg}[1]{\boldsymbol{#1}}
\newcommand{\ox}{\otimes}
\newtheorem{thm}{Theorem}

\newtheorem{defn}{Definition}
\newtheorem{lem}{Lemma}

\newcommand{\id}{\mathbbm{1}}

\nc{\MM}{\widetilde{\M}}
\nc{\Ml}{\M^{\leq}}

\nc{\mleq}{\preceq}
\nc{\mgeq}{\succeq}

\nc{\wt}{\widetilde}

\nc{\SDP}{\text{\rm SDP}}

\nc{\cc}{{\circ\circ}}
\nc{\mnorm}[1]{\norm{#1}{[m]}}

\nc{\F}{\mathcal{F}}
\nc{\M}{\mathcal{M}}

\let\oldproofname\proofname
\rnc{\proofname}{\rm\bf{\oldproofname}}
\rnc{\qedsymbol}{{\color{gray!50!black}\rule{0.6em}{0.6em}}}

\newcommand{\bb}{\begin{equation}}
\newcommand{\bbb}{\begin{equation*}}
\newcommand{\ee}{\end{equation}}
\newcommand{\eee}{\end{equation*}}

\nc{\note}[1]{{\color{blue!90!black} #1}}

\usepackage{wasysym}
\newcommand{\fq}{\bullet}
\newcommand{\mqfi}{M_\text{PE}^{\mc{F}}}
\newcommand{\pospart}[1]{\left[ #1 \right]^+}
\newcommand{\expect}[1]{\mathinner{\langle #1\rangle}}

\makeatletter
\newtheorem*{rep@theorem}{\rep@title}
\newcommand{\newreptheorem}[2]{%
\newenvironment{rep#1}[1]{%
 \def\rep@title{#2 \ref{##1}}%
 \begin{rep@theorem}}%
 {\end{rep@theorem}}}
\makeatother

\newreptheorem{theorem}{Theorem}

\begin{document}
\title{Entanglement between identical particles is a useful and consistent resource}
\author{Benjamin Morris}
\thanks{The contributions of these two authors are impossible to distinguish}
\thanks{\\\href{mailto:benjamin.morris@nottingham.ac.uk}{benjamin.morris@nottingham.ac.uk}}
\thanks{\\\href{mailto:benjamin.yadin@nottingham.ac.uk}{benjamin.yadin@nottingham.ac.uk}}
\thanks{\\\href{mailto:gerardo.adesso@nottingham.ac.uk}{gerardo.adesso@nottingham.ac.uk}}
    \affiliation{School of Mathematical Sciences and Centre for the Mathematics and Theoretical Physics of Quantum Non-Equilibrium Systems,
University of Nottingham, University Park, Nottingham NG7 2RD, United Kingdom}
\author{Benjamin Yadin}
\thanks{The contributions of these two authors are impossible to distinguish}
\thanks{\\\href{mailto:benjamin.morris@nottingham.ac.uk}{benjamin.morris@nottingham.ac.uk}}
\thanks{\\\href{mailto:benjamin.yadin@nottingham.ac.uk}{benjamin.yadin@nottingham.ac.uk}}
\thanks{\\\href{mailto:gerardo.adesso@nottingham.ac.uk}{gerardo.adesso@nottingham.ac.uk}}
    \affiliation{School of Mathematical Sciences and Centre for the Mathematics and Theoretical Physics of Quantum Non-Equilibrium Systems,
University of Nottingham, University Park, Nottingham NG7 2RD, United Kingdom}
    \affiliation{Wolfson College, University of Oxford, Linton Road, Oxford OX2 6UD, United Kingdom}

\author{Matteo Fadel}
\affiliation{Department of Physics and Swiss Nanoscience Institute, University of Basel, Klingelbergstrasse 82, 4056 Basel, Switzerland}
	\affiliation{State Key Laboratory for Mesoscopic Physics, School of Physics \& Collaborative Innovation Center of Quantum Matter,
Peking University, 100871 Beijing, China}

\author{Tilman Zibold}
\affiliation{Department of Physics and Swiss Nanoscience Institute, University of Basel, Klingelbergstrasse 82, 4056 Basel, Switzerland}

\author{Philipp Treutlein}
\affiliation{Department of Physics and Swiss Nanoscience Institute, University of Basel, Klingelbergstrasse 82, 4056 Basel, Switzerland}

\author{Gerardo Adesso}
    \affiliation{School of Mathematical Sciences and Centre for the Mathematics and Theoretical Physics of Quantum Non-Equilibrium Systems,
University of Nottingham, University Park, Nottingham NG7 2RD, United Kingdom}

\begin{abstract}
The existence of fundamentally identical particles represents a foundational distinction between classical and quantum mechanics. Due to their exchange symmetry, identical particles can appear to be entangled –- another uniquely quantum phenomenon with far-reaching practical implications. However, a long-standing debate has questioned whether identical particle entanglement is physical or merely a mathematical artefact. {
In this work, we provide such particle entanglement  with a consistent theoretical description as a quantum resource in processes frequently encountered in optical and cold atomic systems. This leads to a plethora of applications of immediate practical impact. On one hand, we show that the metrological advantage for estimating phase shifts in systems of identical bosons amounts to a  measure of their particle entanglement, with a clearcut operational meaning. On the other hand, we demonstrate in general terms that particle entanglement is the property resulting in directly usable mode entanglement when distributed to separated parties, with particle conservation laws in play. Application of our tools to an experimental implementation with Bose-Einstein condensates leads to the first quantitative estimation of identical particle entanglement. Further connections are revealed between particle entanglement and other resources such as optical nonclassicality and quantum coherence. Overall, this work marks a resolutive step in the ongoing debate by delivering a unifying conceptual and practical understanding of entanglement between identical particles.}
\end{abstract}
\maketitle
%%%%%%%%%%%%%%%%%%%%%%%%%%%%%%%%%%%%%%%%%%%%%%%%%%%%%%%%%%%%%%%%%%%%%%%%%%%%%%%%%%%%%%%%%%%%%%%%%%%%%%%%%%
\section{Introduction}
Identical particles in quantum mechanics have a character quite distinct from those in classical mechanics. Classically, indistinguishability comes from limited abilities of the experimenter; in the quantum world, two particles of the same type, such as electrons, are fundamentally indistinguishable \cite{feynman2011feynman,tuckerman2010statistical}. This feature applies not only to fundamental particles but is also crucial in describing identical composite particle systems such as Bose-Einstein condensates (BECs) \cite{anderson1995observation}. Notably, exchanging two identical quantum particles results in an overall phase change in the wavefunction: no change for bosons and a minus sign for fermions.

These exchange statistics require a symmetric or anti-symmetric wavefunction in the first-quantised formalism. For example, let us denote by $\ket{n_0, n_1}$ a state of identical bosons in which $n_0,\,n_1$ particles have the internal state $\ket{0},\,\ket{1}$ respectively. In the first-quantised picture, we represent $\ket{1,1}$ not as a two-mode state but a symmetric two-particle state
\begin{equation} \label{eqn:fe_example}
	\frac{{\ket{0}}_1 {\ket{1}}_2 + {\ket{1}}_1 {\ket{0}}_2}{\sqrt{2}},
\end{equation}
in which we have attached the fictional labels $1,2$ to the particles. { Formally, the state (\ref{eqn:fe_example}) is {\em entangled}.
However, it may be argued \cite{eckert2002quantum,ghirardi2002entanglement,ghirardi2004general,tichy2011essential,tichy2013entanglement,Dalton2017Quantum} that this ``entanglement'' is unphysical -- since the particles are identical, the labels $1,2$ are meaningless as it is impossible to say which particle has which label. Throughout this work we will  refer to this manifestation of correlations due to exchange symmetry as \emph{Particle Entanglement} (PE) \footnote{Not to be confused with particle entanglement as named in \cite{vaccaro2003entanglement}.}.}

A consensus on the nature of this entanglement has so far been out of reach \cite{ghirardi1977some,pavskauskas2001quantum, shi2003quantum,barnum2004subsystem, zanardi2004quantum, barnum2005generalization,cavalcanti2007useful,ichikawa2008exchange,wei2010exchange,sasaki2011entanglement,benatti2011entanglement, bruschi2012particle, balachandran2013entanglement,benatti2014entanglement,reusch2015entanglement,benatti2017remarks,hyllus2012entanglement}. Some authors view PE as a failure of the mathematical formalism and argue that it should be disregarded in favour of other definitions of identical-particle entanglement \cite{pavskauskas2001quantum,eckert2002quantum,ghirardi2002entanglement,shi2003quantum,barnum2004subsystem,zanardi2004quantum,barnum2005generalization,sasaki2011entanglement,tichy2013entanglement, reusch2015entanglement}. One class of approaches requires talking only about correlations between observables
\cite{barnum2004subsystem, zanardi2004quantum,barnum2005generalization, sasaki2011entanglement,balachandran2013entanglement, reusch2015entanglement}; other authors pursue entirely new definitions of entanglement tailored to the identical-particle setting \cite{ghirardi1977some,pavskauskas2001quantum, ghirardi2002entanglement,  eckert2002quantum,shi2003quantum,ghirardi2004general}. {Many of these approaches are summarised in a recent review \cite{benatti2020entanglement}.}

In order to determine whether there is any meaningful interpretation of PE \emph{per se} we follow the modern approach to entanglement within quantum information theory \cite{horodecki2009quantum}. Here, entangled states are defined as those which cannot be prepared by two or more separated parties who are unable to send quantum information, and are as such limited to local operations (within their own laboratories) and classical communication -- abbreviated as LOCC. Entanglement is then regarded as a \emph{resource} for parties operating under such constraints, and can enable them to perform better at a vast range of tasks including quantum communication \cite{bennett1996mixed}, computation \cite{lloyd1993potentially}, key distribution \cite{shor2000simple}, and metrology \cite{giovannetti2006quantum}, to name a few.

In systems of identical particles, the usable entanglement is that between \emph{modes} \cite{wiseman2003entanglement,schuch2004nonlocal,jones2006entanglement, franco2018indistinguishability, franco2016quantum, vaccaro2003entanglement, compagno2018dealing, Killoran2014Extracting,Dalton2017Quantum, dalton2014new}. This is because (orthogonal) modes are by definition distinguishable systems and so can be addressed individually. Note that these modes need not be spatially separated; we only require that there exist some degree of freedom (such as momentum or internal spin) via which they can be separately addressed. Mode entanglement is distinct from entanglement between particles. For instance, a single particle existing in a superposition of two locations can be viewed as an entangled state of two spatial modes -- but this state clearly contains no PE since there is only one particle. So if mode entanglement is the operationally useful quantity, and is not directly related to PE, why are we interested in the latter? There are strong reasons to believe that PE is a property worth quantifying and may be a resource in certain scenarios. For instance, many-body entangled states of cold atoms, such as spin-squeezed states, can increase precision in metrology thanks to their PE \cite{Braun2018Quantum,pezze2018quantum,Strobel2014Fisher,riedel2010atom,gross2010nonlinear,garbe2019metrological}.

In order to justify PE as a resource, one needs to provide the appropriate setting -- what is the analogue of LOCC for indistinguishable particles? In this work, we first answer that question by finding a physically relevant set of quantum operations in which PE cannot be created. These operations are constructed from combinations of appending vacuum states, performing passive linear unitaries and making either non-demolition measurements of total particle number, or else arbitrary but destructive measurements. We prove that each of these sets of elements is as general as possible while resulting in a consistent theory. In particular, the set of unitaries is physically motivated as ``easy" in many settings, corresponding to beam splitters and phase shifters in optics, and to number-conserving non-interacting hamiltonians in condensed matter systems. These operations, which we call \emph{particle-separable}, define the basis of a \emph{resource theory} for PE. Such an approach has been widely employed recently to pin down a variety of quantum properties beyond entanglement, such as quantum thermodynamics \cite{brandao2013resource}, quantum coherence \cite{streltsov2017colloquium} and asymmetry \cite{bartlett2003classical}. 
With this structure in place, one can begin to rigorously quantify PE and lay the ground for its systematic utilisation in practical tasks.

As a first application, {in Section~\ref{sec:metrology}} we consider the metrological value of PE, in the context of sensing rotations around a collective spin observable. It is known that PE can result in a greater quantum Fisher information, a key figure of merit for the estimation precision achievable with a given state \cite{Hyllus2012Fisher,Toth2012Multipartite}. Beyond just acting as a witness for PE, we show that the enhancement in QFI, suitably quantified, is a monotone under particle-separable operations. It thus follows that operations with particle-entangling power are needed to increase the utility of a state for metrology. This provides a fundamental quantitative assessment of the power of PE as a resource in quantum metrology tasks.

{In Section~\ref{sec:activate}} we use our framework to find the complete setting in which PE is a resource for generating useful mode entanglement between parties. This fully generalises earlier observations by Yurke and Stoler~\cite{Yurke1992Bell} and more recently by Killoran et al.~\cite{Killoran2014Extracting}, the latter providing a starting impetus for this work. Specifically, by ``useful" mode entanglement we mean that which is accessible to parties who are constrained not only by LOCC but also by a local particle-number superselection rule \cite{Wick1952Intrinsic}. The latter constraint renders superpositions of different particle numbers unobservable, and applies when particle number is conserved and the two parties do not have access to a shared phase reference \cite{Bartlett2007Reference}. {
In practical terms, this corresponds to the inability to share laser light with a stable relative phase (in optics) or to share a coherently delocalised BEC (with cold atoms).} Under this limitation, less entanglement can be utilised \cite{schuch2004nonlocal,vaccaro2003entanglement}. We show that useful entanglement can be generated from an initial state by a particle-separable operation exactly when the initial state contains non-zero PE. Furthermore, we find quantitative relations between the amount of input PE and the output useful entanglement. This shows that PE mirrors other quantum resources which may be similarly ``activated" into useful entanglement \cite{Piani2011All,streltsov2015measuring,Ma2016Converting}.
{These results provide a full generalisation of the observations in~\cite{Killoran2014Extracting}. There, it was found that the Schmidt coefficients of a pure PE state remain invariant during its activation into a useful entangled state under a specific class of unitary operations involving non-polarising beam splitters. Thus we have explored the full resource-theoretic meaning of this activation, for the most general states and operations, and quantified it via large classes of entanglement measures.} 

Our results have direct applications to real systems of indistinguishable bosons, in particular entangled states of BECs \cite{simon2002natural,hyllus2012entanglement}. {In Section~\ref{sec:experiment}} we analyse one of a set of recent experimental advances witnessing mode entanglement in BECs \cite{fadel2018spatial,kunkel2018spatially,lange2018entanglement}. We show that these fit into our framework and implement the above resource conversion. {
In particular, our results enable for the first time a quantitative determination of the PE content of the states produced in the experiment, based on quantifiers validated within our resource theory framework.}

Finally, in {Section~\ref{sec:nonclassicality}} we find novel and surprising connections between PE and non-classicality as employed in quantum optics. In that context, classical states are probabilistic mixtures of coherent states \cite{Glauber1963Quantum,Sudarshan1963Equivalence}. States lying outside this set  are non-classical, and are essential in many quantum technological applications \cite{lloyd1999quantum}. Aided by a recent resource theory formulation of non-classicality \cite{Gehrke2012Quantification,tan2017quantifying,Yadin2018Operational,Kwon2019Nonclassicality}, several parallels can be formed between the two disparate topics. 
We find non-classicality to be a necessary but not sufficient prerequisite for PE -- however, non-classicality can be ``unlocked" by using multiple copies of a state. Thus we have a remarkable link between two uniquely quantum resources.

%%%%%%%%%%%%%%%%%%%%%%%%%%%%%%%%%%%%%%%%%%%%%%%%%%%%%%%%%%%%%%%%%%%%%%%%%%%%%%%%%%%%%%%%%%%%%%%%%%%%%%%%%%
\section{Particle identity and superselection rules}
We work with bosonic systems, for which $m$ orthogonal modes have associated annihilation and creation operators $a_i,a_i^\dagger,\, i=0,\dots,m-1$, satisfying the canonical commutation relations $[a_i,a_j]=0,\, [a_i,a_j^\dagger] = \delta_{i,j}$. For a particular choice of modes, the second quantised description is given in terms of the occupation numbers $n_i$ of each mode: $\ket{n_0,\dots,n_{m-1}} \propto (a_{m-1}^\dagger)^{n_{m-1}} \dots (a_0^\dagger)^{n_0} \ket{0,\dots,0}$. All bosonic states then live in the Fock space spanned by such vectors.

In order to make statements about entanglement between particles, it is necessary to ensure that it is even sensible to talk about the particles comprising a state. Such statements are meaningless when a state contains a superposition of different particle numbers. Therefore we permit ourselves only to describe states of definite total particle number  \footnote{An alternative case can be made: a number superselection rule on operations is often in effect in cold atoms and optics. Then a state $\rho_S$ of system $S$ is operationally equivalent to the dephased state $\Phi_S(\rho_S)$, unless one has access to a phase reference $R$ such as a BEC or laser. But appending an additional system can generally contribute to PE (Appendix~\ref{app:appending}), so $R$ must be included within the description as a resource. The joint system $SR$ is then described as diagonal in total number.} -- or probabilistic mixtures of such states \cite{wiseman2003entanglement,Dalton2017Quantum}. Mathematically, this is described by a particle-number \emph{superselection rule} (SSR), which forces any state $\rho$ under consideration to be block-diagonal with respect to the total number operator $\hat{N}$, also expressed as $[\rho,\hat{N}]=0$. (We distinguish between the operator $\hat{N}$ and its eigenvalues $N$.) Similarly, all considered operations $\mc{E}$ (i.e., completely positive maps on the set of states) must respect the SSR. This is ensured by taking only \emph{covariant} operations, defined by commutation $[\mc{E}, \mc{U}_\theta]=0$ with the phase rotation channel $\mc{U}_\theta(\rho) = e^{-i\theta \hat{N}} \rho e^{i\theta \hat{N}}$ for all $\theta$ \cite{Bartlett2007Reference}. Equivalently, covariant operations can be performed via a dilation involving an initially number-diagonal environment and a global particle number conserving unitary interaction \cite{Keyl1999Optimal}.

Any state of definite particle number $N=\sum_i n_i$ can be written in the first quantised picture, where each particle has an internal state in the single-particle space $\mc{H}_1$ of dimension $m$ (so that there is one degree of freedom for each mode). The overall state then lies in the symmetric subspace of the $N$-system space, denoted by $\mc{H}_N = \mc{S}[ \mc{H}_1^{\ox N}]$. A general mixture of particle numbers $\rho = \sum_N p_N \rho^{(N)}$ can be described as being a state on $\mc{S}[\mc{H}_1^{\ox N}]$ with probability $p_N$. Where necessary, we distinguish between the first and second quantised forms of a pure state using the notation ${\ket{\psi}}^{\fq}$ and ${\ket{\psi}}$ respectively, and similarly $\rho^{\fq}$ and $\rho$ for a mixed state.  \\

%%%%%%%%%%%%%%%%%%%%%%%%%%%%%%%%%%%%%%%%%%%%%
\section{PE as a resource}
 A resource theory is defined by two components: the set of \emph{free states} $\mc{S}$, which possess no resource, and the set of \emph{free operations} $\mc{O}$, which do not add any new resource into the system. (One also tends to think of free operations as possible to perform without any resource, although this interpretation is not always clear.)

The set of free states for PE is straightforward to define. For fixed particle number $N$, they must be non-entangled (separable) states in the first-quantised picture. Due to symmetry, a pure $N$-particle free state is thus of the form ${\ket{\Psi}^{\fq}} = {\ket{\psi}}^{\ox N}$, also known as a coherent spin state \cite{Giraud2010Quantifying,pezze2018quantum}. In second-quantised form, we have $\ket{\Psi} \propto (c_\psi^\dagger)^N \ket{0}$, where $c^\dagger_\psi = \sum_i \psi_i a^\dagger_i$ creates a single particle in an arbitrary mode $\psi$. A mixed $N$-particle free state is by definition symmetric and separable -- it turns out (see Appendix \ref{app:free_states}) that this is equivalent to the form
\begin{equation} \label{eqn:separable}
	\rho^\fq = \sum_i \lambda_i {\proj{\psi_i}}^{\ox N}, \; \lambda_i \geq 0, \; \sum_i \lambda_i=1.
\end{equation}
Then the full set of free states -- which we name \emph{particle-separable} -- consists of those $\rho = \sum_N p_N \rho^{(N)}$ such that each of these components in the first-quantised picture is of the form (\ref{eqn:separable}).

We may then choose as free operations any set that preserves particle-separability. This is required in order to ensure a consistent notion of a resource. There is often tension between the desire for mathematical generality of these operations and wanting them to have a known physical implementation. In our approach, we do not take the largest set of quantum operations preserving particle-separability, but instead construct a physically transparent set from elementary types of operations. We prove that each of these elements is as general as possible.

In the spirit of the Stinespring dilation for quantum operations \cite{nielsen2010quantum}, we construct our free operations out of three basic steps: (i) appending ancilliary modes; (ii) global unitary operations; (iii) projective measurements. We investigate each of these in turn.

\emph{(i) Appending ancilliary modes:~}In mathematical terms, the action of appending to a state $\rho$ another set of modes in a fixed state $\sigma$ means $\rho \to \rho \ox \sigma$ in second quantisation. In order to consider this a free operation, we restrict $\sigma \in \mc{S}$. In most resource theories this operation would preserve the set of free states \cite{chitambar2019quantum}. However, the present theory is unusual in that this generally fails -- the simplest example is appending the single-particle state $\ket{1}$ to another copy of itself, as $\ket{1,1} \equiv \ket{1}\ket{1}$ is not particle-separable. The reason for this is that appending particles in new modes requires symmetrisation in the first quantised picture, which creates PE. As we show in Appendix \ref{app:appending}, the only ancilliary state $\sigma$ that guarantees preservation of free states is the vacuum.

\emph{(ii) Unitaries:~}The covariance condition for unitaries means that they preserve particle number: $[U,\hat{N}]=0$. Consider first the component $U^{(N)}$ acting on the $N$-particle subspace. We see that $U^{(N)}$ preserves $\mc{S}$ if and only if it has the first-quantised action $U^{(N)\fq} {\ket{\psi}}^{\ox N} = {\ket{\phi}}^{\ox N}$ for every $\ket{\psi} \in \mc{H}_1$, where $\ket{\phi}$ can depend on $\ket{\psi}$. Perhaps unsurprisingly, this is equivalent to $U^{(N)\fq} = u^{\ox N}$ for any single-particle unitary $u$, although the argument is not immediate and invokes Wigner's theorem on inner-product-preserving transformations \cite{wignerstuff} (see Appendix \ref{app:unitaries}). In principle, this $u$ could be different for each number $N$ -- however, the introduction of number measurements below implies that we lose no generality by taking a fixed $u$. Such unitaries have a simple second-quantised description via their action on ladder operators: $U^\dagger a_i^\dagger U = \sum_j u_{ij} a_j^\dagger$, where $u_{ij}$ are the elements of a unitary matrix. They describe single-particle rotations without interaction, acting identically on all particles, and correspond to passive linear operations in optics, which are easily generated by beam splitters and phase shifters \cite{Reck1994Experimental}.

\emph{(iii) Projective measurements:~} A projective measurement is given by a set of projectors $\Pi_i$ which are orthogonal and complete: $\Pi_i \Pi_j = \delta_{i,j} \Pi_i,\, \sum_i \Pi_i = \id$. As for unitary operations, these must adhere to the SSR, $[\Pi_i, \hat{N}] = 0$, and preserve the set of particle-separable states, $\Pi_i^{\fq (N)} {\ket{\psi}}^{\ox N} \propto {\ket{\phi}}^{\ox N}$. However, we find that these conditions are only met by a measurement of total particle number (see Appendix \ref{app:measurements}). In order to enlarge the set of available measurements, we allow \textit{destructive measurements}, in which the measured modes are subsequently discarded. In Appendix \ref{app:measurements} we demonstrate that this relaxation allows any measurement adhering to the SSR to be performed on the system without introducing PE. Such destructive measurements correspond to the majority of experimental photon- and atom-counting techniques.

The set $\mc{O}$ of \emph{particle-separable operations} is defined as all possible protocols which result from combinations of the above elements, including possible conditioning of future operations on the results of measurement outcomes. We also allow for the use of classical randomness and coarse-graining -- i.e., forgetting measurement outcomes. Mathematically, an element in $\mc{O}$ is represented as a quantum instrument, which is a set of CP maps $\mc{E}_i$ where each $i$ labels a single (possibly coarse-grained) measurement outcome and the sum $\sum_i \mc{E}_i$ is deterministic (trace-preserving). Note that an instrument can equivalently be represented as a deterministic channel $\mc{F}(\rho) = \sum_i \mc{E}_i(\rho) \ox {\proj{i}}_X$, where the outcome is stored in a classical system $X$ \cite{adesso2016measures}.

With this structure in place, we can now move naturally to define measures $M_{\text{PE}}$ of PE. As is standard in quantum resource theories \cite{chitambar2019quantum}, we require that any measure of PE fulfills the following three conditions. \emph{Condition (i)}--It must not detect PE when there is none, meaning $M_{\text{PE}}(\rho)=0$ for all $\rho \in \mc{S}$ (and optionally the converse may be required). \emph{Condition (ii)}--$M_{\text{PE}}$ must be a \emph{monotone}, i.e.~cannot increase under the action of any particle-separable operation. This reflects the idea that particle-separable operations cannot inject additional PE into the system. Monotonicity can be stated either deterministically, $M_{\text{PE}}(\rho) \geq M_{\text{PE}}(\mc{E}[\rho])$ for any channel $\mc{E} \in \mc{O}$, or probabilistically, $M_{\text{PE}}(\rho) \geq \sum_i p_i M_{\text{PE}}(\rho_i)$ for an instrument $\{\mc{E}_i\}$ in $\mc{O}$ with outcomes $p_i \rho_i = \mc{E}_i(\rho)$. \emph{Condition (iii)}--Convexity, i.e., being non-increasing under probabilistically mixing different states, $\sum_i p_i M_\text{PE}(\rho_i) \geq M_\text{PE}(\sum_i p_i \rho_i)$.

A straightforward class of PE measures are given by the minimal distance between a state and the set of particle-separable states:
\begin{equation} \label{eqn:distance_measure}
    M_\text{PE}^D(\rho) := \min_{\sigma \in \mc{S}} D(\rho,\sigma),
\end{equation}
where $D$ is any suitable measure of distinguishability between two quantum states. {
Conditions (i,iii) and the deterministic version of (ii) are met whenever $D$ is contractive under quantum channels (so that $D(\mc{E}(\rho),\mc{E}(\sigma)) \leq D(\rho,\sigma)$ for any channel $\mc{E}$) and jointly convex in its arguments; other properties may guarantee ensemble monotonicity (ii) (see Appendix~\ref{app:measures} and Ref.~\cite{chitambar2019quantum}).} \\

{
\section{Quantifying metrological power of PE} \label{sec:metrology}
Now that we have determined the set of protocols under which PE may abstractly be considered a resource, we are in a position to demonstrate concrete tasks in which it is useful. In this section, we use our resource theory to demonstrate a quantitative connection between PE and quantum metrology. A typical metrological setting involves a parameter $\theta$ encoded into a system, such that the experimenter is given one of a parameterised family of states $\rho_\theta$, and the task is to estimate $\theta$ via measurements. Here, we focus on the case of unitary encoding, whereby an initial state $\rho$ evolves under a given Hamiltonian $H$, so that $\rho_\theta = e^{-i\theta H}\rho e^{i\theta H}$. An important figure of merit is the quantum Fisher information (QFI) $\mc{F}(\rho,H) := -4\partial_\theta^2 \mathrm{Fid}(\rho,\rho_\theta)|_{\theta=0}$, where $\mathrm{Fid}(\rho,\sigma) = \Tr \sqrt{\sqrt{\rho}\sigma \sqrt{\rho}}$ is the fidelity between two states. The QFI can be thought of as a measure of speed of evolution for $\rho_\theta$ under the dynamics generated by $H$. Its importance for metrology is given by the (quantum) Cram\'er-Rao bound, which says that the uncertainty $\Delta \theta$ in estimating $\theta$ is lower-bounded by $(\Delta \theta)^2 \geq 1/(n \mc{F}(\rho,H))$ with $n$ copies of $\rho_\theta$ provided \cite{Paris2009Quantum}.

PE is known to be a necessary resource for a quantum-enhanced metrology \cite{Braun2018Quantum,pezze2018quantum}. For $N$ qubits, one can define total spin components $\boldsymbol{S^\alpha} := \sum_{i=1}^N \sigma^\alpha_i/\sqrt{N},\, \alpha=x,y,z$, where $\sigma^\alpha_i$ is a Pauli matrix acting on the $i$th particle; the spin in any direction $\boldsymbol{n}=(n^x,n^y,n^z)$, with $\abs{\boldsymbol{n}}=1$, is denoted as $\boldsymbol{n}\cdot \boldsymbol{S}$. Then, for any particle-separable state, we have $\mc{F}(\rho,\boldsymbol{n}\cdot\boldsymbol{S}) \leq 1$ \cite{Toth2012Multipartite,Hyllus2012Fisher}. Exceeding this bound witnesses PE, with the maximum possible QFI being $N$. A tighter bound, applicable to any Hamiltonian of the form $H = \sum_{i=1}^N h_i/\sqrt{N}$, was more recently proven \cite{Gessner2016Resolution}:
\begin{equation}
    \rho \in \mc{S} \Rightarrow \mc{F}(\rho,H) \leq 4 \sum_{i=1}^N V\left(\rho,\frac{h_i}{\sqrt{N}}\right) = 4V(\rho,h_1).
\end{equation}
Based on this inequality, we define the following quantity as the amount by which the QFI exceeds the limit for particle-separable states:
\begin{equation}
	\mqfi(\rho) := \max_{h: \|h\| = 1} \pospart{ \mc{F}(\rho,H) - 4V(\rho, h) },
\end{equation}
where $H = \bigoplus_N H^{(N)},\, H^{(N)\fq} = \sum_{i=1}^N h_i/\sqrt{N}$, $\pospart{x} = \max \{x,0\}$ denotes the positive part of $x$,  and the maximisation is performed over all single-particle observables $h$ with unit operator norm. %$\|h\| = \lambda_\text{max}(\abs{h})$. $\pospart{x} = \max \{x,0\}$ denotes the positive part of $x$.
The expectation value of a single-particle operator $h$ in a number-varying state $\rho = \sum_N p_N \rho^{(N)}$ is defined as
\begin{equation}
	\expect{h}_\rho := \sum_N p_N \Tr[ \rho^{(N)\fq} h_1] = \sum_N p_N \frac{1}{N}\Tr\left[ \rho^{(N)\fq} \sum_{i=1}^N h_i \right],
\end{equation}
so that $V(\rho,h) := \expect{h^2}_\rho - \expect{h}_\rho^2$.

We can also extend the measure to include settings where one records measurement outcomes in a classical memory $M$. In this case, a state is in ``quantum-classical" form $\rho_{SM} = \sum_m p_m \rho_{S|m} \ox {\proj{m}}_M$, where $p_m$ is the probability of outcome $m$, $\rho_{S|m}$ the corresponding conditional state of the system $S$, and the states $\{\ket{m}\}$ form an orthonormal basis for the memory $M$. For such a state, the observable $h$ is understood to only act on $S$ and not on the memory $M$, i.e.,
\begin{equation}
	\mqfi(\rho_{SM}) := \max_{h_S: \|h_S\| = 1} \pospart{ \mc{F}(\rho_{SM},H_S) - 4V(\rho_{SM},h_S)}.
\end{equation}
As a consequence of this definition, the QFI part can be expressed as an average over measurement outcomes, $\sum_m p_m \mc{F}(\rho_{S|m},H_S)$, while the variance part is calculated for the whole ensemble $\rho_{SM}$.

Remarkably, we find that $\mqfi$ is not only a witness of PE, but also a monotone under particle-separable operations (without feed-forward):
\begin{thm} \label{thm:qfi_monotone}
    $\mqfi$ is convex and satisfies $\mqfi(\rho) = 0 \; \forall \rho \in \mc{S}$.
    Moreover, let $\mc{E}_{S \to SM} \in \mc{O}$ contain a single measurement round, such that no conditional operations are performed after the measurement. We may write $\mc{E}_{S\to SM}(\rho_S) = \sum_m \mc{E}^m(\rho_S) \ox {\proj{m}}_M$, where $\mc{E}^m$ is the operation applied to $\rho_S$ conditioned on outcome $m$. Then
    \begin{equation}
        \mqfi(\rho_S) \geq \mqfi(\mc{E}_{S \to SM}[\rho]).
    \end{equation}
\end{thm}
The proof is presented in Appendix~\ref{app:qfi}. Note that $\mqfi$  may vanish for some particle-entangled states -- however, for pure states, it does faithfully detect PE \cite{Gessner2016Resolution}. The monotonicity result demonstrates that, beyond being a witness, $\mqfi$ captures the ordering of particle-entangled states under the free operations in the resource theory developed in this paper. From a practical perspective, this shows the limitations on particle-separable operations for enhancing the utility of a state for metrology, and ultimately provides an original and operationally motivated tool to quantify PE by means of its metrological value, in addition to the distance-based measures presented earlier.

A simplification is possible in the special case of two modes (i.e., when the particles are qubits). Given $\|h\|=1$, without loss of generality we can write $h = \proj{0} + \lambda\proj{1}$ in some basis, where $\abs{\lambda} \leq 1$. Since the QFI and variance are invariant under constant shifts of the observable, we can shift $h$ to $h - (\frac{1+\lambda}{2}) I = (\frac{1-\lambda}{2})\sigma^z$, thus getting
\begin{align}
	\pospart{\mc{F}(\rho,H) - 4V(\rho,h)} & = \left(\frac{1-\lambda}{2}\right)^2 \pospart{\mc{F}(\rho,Z) - 4V(\rho,\sigma^z)} \nonumber \\
		& \leq \pospart{\mc{F}(\rho,Z) - 4V(\rho,\sigma^z)},
\end{align}
where $Z^{(N)\fq} = \sum_{i=1}^N \sigma^z_i/\sqrt{N}$. Equality is obtained for $\lambda = -1$, i.e., $h = \sigma^z$. Hence, in this case, the only remaining degree of freedom is the eigenbasis of $h$, which can be translated into a spin direction $\boldsymbol{n}$:
\begin{equation}
	\dim \mc{H}_1 = 2 \Rightarrow \mqfi(\rho) = \max_{\boldsymbol{n}: \abs{\boldsymbol{n}}=1} \pospart{\mc{F}(\rho,\boldsymbol{n} \cdot \boldsymbol{S}) - 4V(\rho,\boldsymbol{n} \cdot \boldsymbol{\sigma})}.
\end{equation}

Note how, in addition to  generalising (and tightening) the QFI witnesses proposed in Refs.~\cite{Toth2012Multipartite,Hyllus2012Fisher}, our measure $\mqfi$ differentiates itself by explicitly including the variance of the single-particle observable, rather than being used to bound the measure. The importance of its inclusion is apparent in the proof of Theorem~\ref{thm:qfi_monotone}, specifically in order to show that $\mqfi$ is invariant under the addition of vacuum modes. When new modes are included, the set of possible $h$ observables increases, allowing for a greater possible QFI -- we may have $\max_{h'}\mc{F}(\rho_S \ox {\proj{0}}_A, H') > \max_{h}\mc{F}(\rho_S,H)$. The variance component nontrivially compensates for this effect.
}

%%%%%%%%%%%%%%%%%%%%%%%%%%%%%%%%%%%%%%%%%%%%%
\section{Activating PE} \label{sec:activate}
{ Here, we describe another important task for which the utility of PE as a resource is manifest. The original seeds of the \emph{activation} protocol that we analyse here are in work by Yurke and Stoler, who noted that two particles produced from separated, independent sources can in fact be used to violate a Bell inequality \cite{Yurke1992Bell}. The protocol that we present is a direct application of our resource theoretic formulation and constitutes a full generalisation of \cite{Killoran2014Extracting}.}

Consider two separated parties, $A$ and $B$, who want to perform some joint quantum information protocol but are constrained to classical communication and additionally lack a shared phase reference (conjugate to the number observable $\hat{N}_A$ or $\hat{N}_B$). { A phase reference would be provided by a shared state containing coherence with respect to the local number observable $\hat{N}_A$ (or $\hat{N}_B$). In optics, a typical example is a laser coherently split into modes held by each party, maintaining a fixed phase relationship. The analogue in cold atoms is a coherently distributed BEC. Extensive discussions of the relationship between SSRs and phase references can be found in Refs.~\cite{Dalton2017Quantum,Bartlett2007Reference}.}

\begin{figure}
	\centering
	\fbox{\includegraphics[scale=0.185]{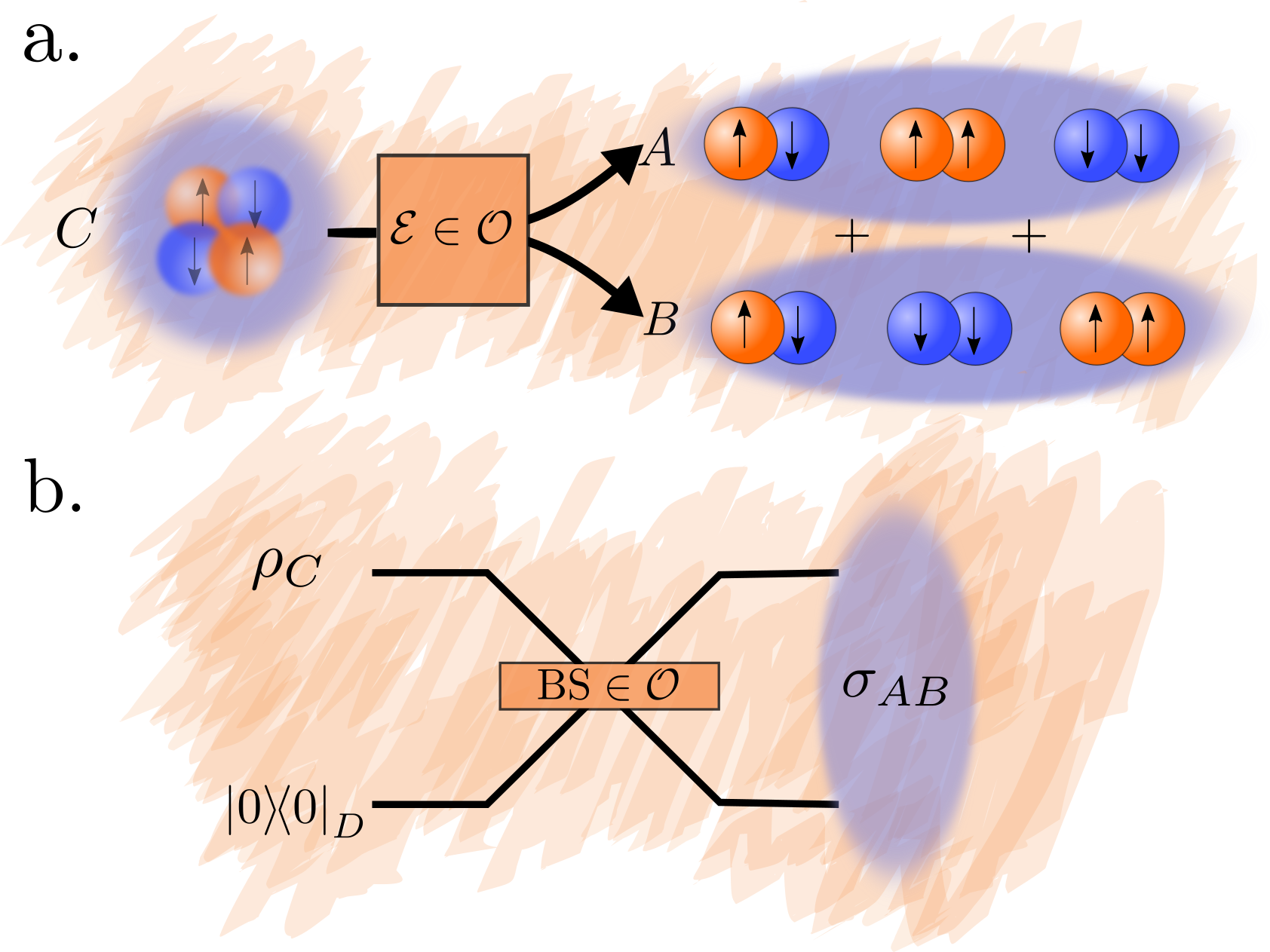}}
	\caption{ a. Conversion protocol between PE and  SSR-entanglement via the quantum operation $\mc{E}\in\mc{O}$. The operation $\mc{E}$ converts a system of identical particles with PE into a bipartite state, whose SSR-entanglement can be extracted and utilised in quantum information tasks. The above diagram depicts the transformation $\ket{2,2}_C\overset{\mc{E}\in\mc{O}}{\longrightarrow}\left(\ket{1,1}_A\ket{1,1}_B+\ket{2,0}_A\ket{0,2}_B+\ket{0,2}_A\ket{2,0}_B\right)$, having post-selected $N_A=N_B=2$.\\ b. An example of a particle-separable operation is the action of a beam-splitter with a vacuum, which can be used to activate the PE present in the state $\rho_C$.  \label{fig:protocol}}
\end{figure}

While each party may be unconstrained in their local operations, without sharing a phase reference, the amount of entanglement accessible to them is reduced by the application of an effective local SSR \cite{Bartlett2007Reference}. This SSR corresponds to both local particle numbers $\hat{N}_A$ and $\hat{N}_B$. A third party $C$ is tasked with providing $A$ and $B$ with a shared entangled state that they can use. To accomplish this, $C$ has an initial resource state $\rho_C$ of $m$ modes and can process it using any particle-separable operation $\mc{E}$ before distributing $m_A$ and $m_B$ modes to each of $A$ and $B$. (Recall that the operation $\mc{E}$ may introduce new vacuum modes and trace out some modes; see Fig.~\ref{fig:protocol}). The question is: how much useful entanglement can be extracted in this way from $\rho_C$?

Let $\sigma_{AB} = \mc{E}(\rho_C)$ be the output state sent to $A$ and $B$, where $\mc{E} \in \mc{O}$ is the distribution operation performed by $C$. (Without loss of generality, using classical flags, we can take this to be deterministic.) Due to the local SSR, from the perspective of $A$ and $B$, this state is operationally as useful as the state $\Phi_A \ox \Phi_B(\sigma_{AB})$ \cite{schuch2004quantum}, where $\Phi_S$ is the dephasing channel local to subsystem $S$, removing quantum coherences between states of differing local number $\hat{N}_S$ \footnote{This may be written equivalently as a phase average $\Phi_S(\rho) = \int_0^{2\pi} \dd \theta \, e^{-i\theta \hat{N}_S} \rho e^{i \theta \hat{N}_S} / 2\pi$ or as a ``measure-and-forget'' operation of the local number: $\Phi_S(\rho) = \sum_n P_{n,S} \rho P_{n,S}$, where $P_{n,S}$ is the projector onto the subspace of $n$ particles in $S$.}.

For any measure $E$ of bipartite entanglement, we can then define the corresponding measure of entanglement accessible to $A$ and $B$ \cite{Bartlett2007Reference}:
\begin{equation} \label{eqn:ssr_entanglement}
	E_\text{SSR}(\sigma_{AB}) := E\left( \Phi_A \ox \Phi_B (\sigma_{AB}) \right) \leq E(\sigma_{AB}).
\end{equation}
We say that a state $\sigma_{AB}$ is \emph{SSR-separable} whenever it has vanishing accessible entanglement -- i.e., when $\Phi_A \ox \Phi_B(\sigma_{AB})$ is separable -- and \emph{SSR-entangled} otherwise. The inequality in (\ref{eqn:ssr_entanglement}) follows from the fact that $\Phi_A \ox \Phi_B$ is a local operation -- the local SSR generally reduces the amount of accessible entanglement. The aspect of the entanglement in $\sigma_{AB}$ that is \emph{inaccessible}, sometimes referred to as ``fluffy bunny entanglement" \cite{wiseman2004ferreting}, is connected with superpositions of local number. Note that Wiseman and Vaccaro \cite{wiseman2003entanglement} proposed the same class of measures \eqref{eqn:ssr_entanglement} and found such SSR-entanglement to require non-zero PE in the case of two particles.

We prove that PE in the initial state $\rho_C$ is precisely the resource enabling the distribution of SSR-entanglement. Our first result is that the mapping between the two types of entanglement is faithful, in that SSR-entanglement can be extracted exactly when there is nonzero PE (see Appendix \ref{app:activating} for the proof):
\begin{thm}\label{thm:activation_faithful}
	There exists an activation operation $\mc{E}_{C \to AB} \in \mc{O}$ creating an SSR-entangled state $\sigma_{AB}$ from $\rho_C$ if and only if $\rho_C \not \in \mc{S}$.
\end{thm}
Moreover, almost any operation of the following type is sufficient to activate PE into non-zero SSR-entanglement: for each mode $i$ in $C$, attach a new mode in the vacuum state, and perform a global passive-linear unitary coupling the modes (as in Fig.~\ref{fig:protocol}b). We say ``almost all" because the unitary must not be trivial by failing to couple some of the modes. {Ref.~\cite{Killoran2014Extracting} examined activation for a specific class of unitary interactions, namely a set of beam-splitters with identical transmission coefficients.} However, we see that a much more general statement is possible, expanding the scope to all particle-separable operations.

Beyond the faithful mapping between nonzero resources, we now quantitatively relate the input and output forms of entanglement. One approach uses measures of both PE and SSR-entanglement constructed in the same way. Recall the distance-based measure of PE $M^D_\text{PE}$; by the same recipe, one can construct a measure of SSR-entanglement (see Appendix \ref{app:activating}):
\begin{align}\label{eqn:e_sr_distance}
	E^D_\text{SSR}(\rho_{AB}) & = E^D(\Phi_A \ox \Phi_B[\rho_{AB}]) \nonumber \\
	    & := \min_{\sigma_{AB} \in \text{ sep.}} D(\Phi_A \ox \Phi_B[\rho_{AB}], \sigma_{AB}).
\end{align}
As shown in Appendix \ref{app:ssr_entanglement}, when $\rho$ respects the local SSR, the minimisation can be equivalently performed over the smaller set of $\sigma_{AB}$ being separable and respecting the local SSR. Using this, we have:

\begin{thm}\label{thm:activation_inequality}
	For any activation $\mc{E}_{C\to AB} \in \mc{O}, \, E^D_\text{SSR}( \mc{E}_{C\to AB}[\rho_C]) \leq M^D_\text{PE}( \rho_C)$.
\end{thm}
This shows that the amount of accessible entanglement extracted never exceeds the initial amount of PE. Note, however, a subtlety: in general, this inequality is strict (apart from when both sides are zero), due to a necessary reduction in entanglement after applying the dephasing operation $\Phi_A \ox \Phi_B$ and removing the ``fluffy bunny entanglement".

Alternatively, we can take any measure of SSR-entanglement and use it to construct a new measure of PE. This is given by the maximal amount of SSR-entanglement which can be created from a certain initial state:
\begin{thm}\label{thm:measure_from_e_ssr_main}
	For any (convex) entanglement measure $E$, the quantity defined as
	\begin{equation}
		M^E_\text{PE}(\rho) := \sup_{\mc{E}_{C \to AB} \in \mc{O}} E_\text{SSR}\left( \mc{E}_{C \to AB}[\rho_C] \right)
	\end{equation}
	is a (convex) measure of PE.
\end{thm}
In other words, for any entanglement measure $E$, the corresponding quantity $M^E_{PE}$ satisfies criteria (i-iii). Theorem~\ref{thm:measure_from_e_ssr_main} gives a precise quantitative version of the statement that PE is the resource for producing SSR-entanglement. \\

\section{Experimentally measuring PE} \label{sec:experiment}
 In this section we demonstrate that our resource theory for describing PE and its activation encompasses recent experimental investigations \cite{fadel2018spatial,kunkel2018spatially,lange2018entanglement} converting PE into useful mode entanglement.
 { This enables us to promptly analyse the experimental data from \cite{fadel2018spatial} in order to extract a lower bound to a measure of PE. To the best of our knowledge, this constitutes the first instance of quantitative estimation of PE in an experiment.}

\begin{figure*}
	\centering
    \fbox{\includegraphics[width=.97\textwidth]{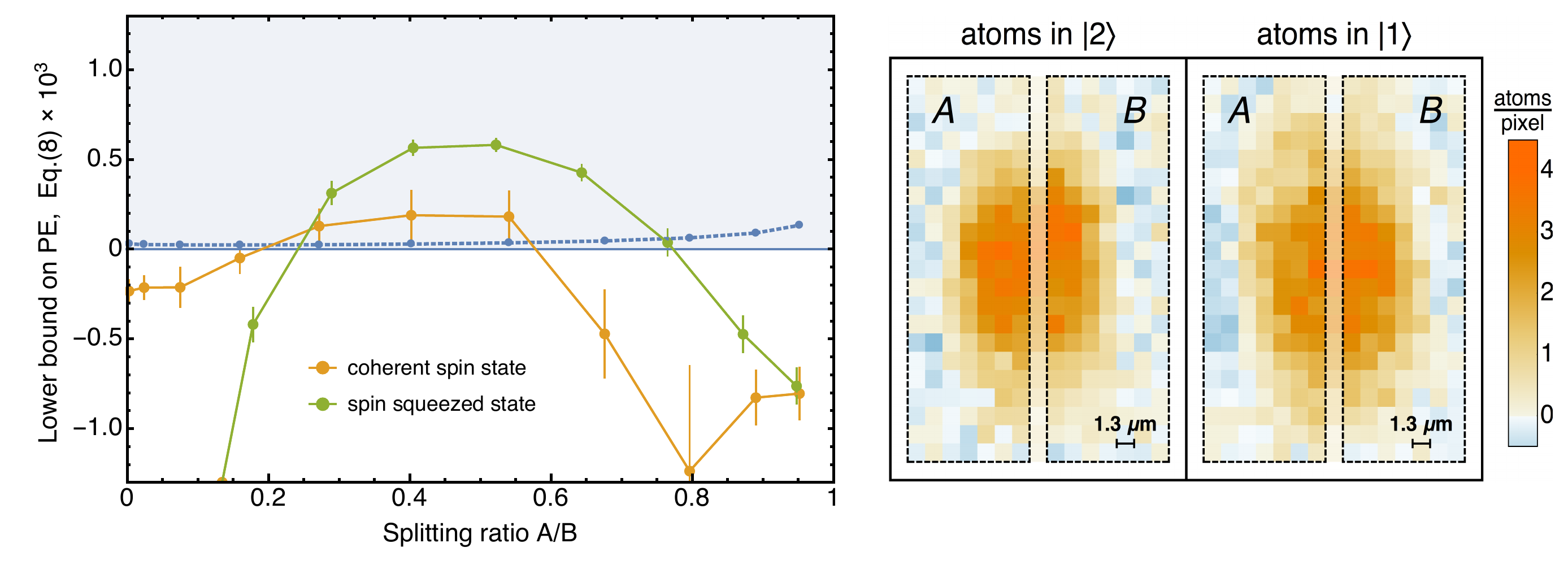}}
	\caption{Based on the measurements \cite{fadel2018spatial} we are able to extract the lower bound given by the right-hand side of \eqref{eqn:entmlowerboundmain}, on the PE measure $M^{\Tr}_{\text{PE}}$. The two sets of points correspond to initialising the BEC either in a spin squeezed state (green), where Particle Entanglement is present, or in a coherent spin state (orange), which is particle-separable. Along the horizontal axis we vary the relative size of the two regions $A$ and $B$ from which we extract the spin values as explained in \cite{fadel2018spatial}. In the experiment, technical limitations in the resolution of assigning the atomic spins to the regions can lead to classical correlations, resulting in apparent entanglement. We give an upper bound for these correlations as the blue dashed line. For intermediate splitting ratios we find significant entanglement in the case of the spin squeezed state while the coherent spin state remains compatible with no particle entanglement within experimental error. On the right we show single-shot absorption images of the atomic densities for the two internal degrees of freedom, with an example of regions $A$ and $B$ used to define the collective spins $\hat{S}^A$ and $\hat{S}^B$ entering in \eqref{eqn:entmlowerboundmain}.  \label{fig:expwitness}}
\end{figure*}

The experimental method is as follows -- see \cite{fadel2018spatial} for more details. The BEC is initialised in a spin-squeezed state, which possesses PE. The BEC is then released from its trap and allowed to expand, and the spin components of the two spatially separated regions are measured. During the expansion, the effect of interactions between atoms on their spin state is negligible such that this step can be regarded as a beam-splitter operation and hence falls within our set of particle-separable operations \footnote{The interaction of ultracold $^{87}\text{Rb}$ atoms depends only very weakly on their spin state. During the expansion of the BEC, the interactions therefore do not affect the spin state and are furthermore quickly rendered small due to the decreasing density \cite{castin1996bose}.}. Furthermore the measurement of spin components of the spatially separated regions adheres to the local SSR \footnote{Due to technical limitations a fraction of the atomic spins in a gap between the two regions is discarded in the measurement process.}. The correlations between the two spatial regions are held in the spin components of the condensate atoms. In particular the $z$-component of the spin in regions $A,B$ is defined as $\hat{S}^{(A,B)}_{z}:=\frac{1}{2\eta^{(A,B)}_{\text{eff}}}\left(\hat{N}_1^{(A,B)}-\hat{N}_2^{(A,B)}\right)$ where $1,2$ correspond to the internal degree of freedom of the atom and $\eta^{(A,B)}_{\text{eff}}$ accounts for finite spatial resolution in the detection of the BEC. Other spin components, e.g. $\hat{S}^{(A,B)}_x$ and $\hat{S}^{(A,B)}_y$, can be measured by applying appropriate spin rotations before detection, these local rotations also being allowed within SSR constraints.

In Ref.~\cite{fadel2018spatial} the authors showed how these local spin measurements can violate the inequality \cite{giovannetti2003characterizing}
    \begin{align}\label{cond:sepMain}
        \frac{4\text{Var}\left( g_z \hat{S}_z^A+\hat{S}_z^B\right)\text{Var}\left( g_y \hat{S}_y^A+\hat{S}_y^B\right)}{\left(\left|g_z g_y\right|\left|\expval{\hat{S}_x^A}\right|+\left|\expval{\hat{S}_x^B}\right|\right)^2}\geq 1 \;,
\end{align}
in terms of variances and average values of spin observables. The condition~\eqref{cond:sepMain} is satisfied by all separable states and for any real constants $g_{y,z}$, therefore certifying entanglement between system $A$ and $B$ whenever a violation is measured.

In Appendix \ref{app:experiment}, we linearise \eqref{cond:sepMain} and use Theorem \ref{thm:measure_from_e_ssr_main} to derive a lower bound on a measure of PE:
\begin{align}\label{eqn:entmlowerboundmain}
    M_\text{PE}^{\Tr}(\rho) \geq & \frac{-1}{\mc{N}} \left[ \text{Var}\left( g_z \hat{S}_z^A+\hat{S}_z^B\right)_\rho +\text{Var}\left( g_y \hat{S}_y^A+\hat{S}_y^B\right)_\rho \right. \nonumber \\
        & \left. \quad - \expval{\left|g_z g_y\right| \hat{S}_x^A+\hat{S}_x^B}_\rho \right], \nonumber \\
    \mc{N} & := \frac{1}{4}\left(\frac{\abs{g_z}N_1^A}{\eta^{A}_{\text{eff}}}+\frac{N_1^B}{\eta^{B}_{\text{eff}}}\right)^2+\frac{1}{4}\left(\frac{\abs{g_y}N_1^A}{\eta^{A}_{\text{eff}}}+\frac{N_1^B}{\eta^{B}_{\text{eff}}}\right)^2  \nonumber \\
        & \quad + \left(\frac{|g_z g_y|N_1^A}{\eta^{A}_{\text{eff}}}+\frac{N_1^B}{\eta^{B}_{\text{eff}}}\right),
\end{align}
where $M_\text{PE}^{\Tr}$ is defined according to \eqref{eqn:distance_measure} with the trace distance $D_{\Tr} (\rho,\sigma) := \frac{1}{2} \Tr \abs{\rho-\sigma}$. We show an evaluation of this bound using experimental results in Fig.~\ref{fig:expwitness}. The parameters $g_{y,z}$ are optimised numerically so that the left-hand side of \eqref{cond:sepMain} is minimised, as this expression is more robust than \eqref{eqn:entmlowerboundmain} against experimental noise. This plot clearly shows a positive amount of PE has been activated from a spin squeezed BEC and none from a coherent spin BEC state, as predicted from our theory.

{ The case study presented in this section reveals how our resource theoretic characterisation of PE unlocks useful quantitative tools that can be readily employed by the cold atoms community to benchmark present and future experiments, including demonstrations of entanglement production and manipulation, sensing and metrology tasks, and other quantum technology protocols empowered by PE.}
\\

%%%%%%%%%%%%%%%%%%%%%%%%%%%%%%%%%%%%%%%%%%%%%
\section{Connections to non-classicality} \label{sec:nonclassicality}
While coherent spin states are considered classical in cold atoms settings with fixed particle number, continuous-variable coherent states in quantum optics provide the model of classical light. Non-classical states display features such as photon anti-bunching, sub-poissonian statistics and squeezing \cite{Gerry2004Introductory}, and form the basis of many quantum technological applications \cite{lloyd1999quantum} As has been recently appreciated, \cite{Gehrke2012Quantification,Yadin2018Operational,Kwon2019Nonclassicality} non-classicality can also be quantified with its own resource theory. In this section we demonstrate some remarkable connections between the resources theories for PE and non-classicality.

Recall that a single-mode coherent state $\ket{\alpha}$ is an eigenstate of the annihilation operator: $a \ket{\alpha} = \alpha \ket{\alpha}$, and a multi-mode coherent state may be written as $\ket{\bg{\alpha}} := \ket{\alpha_1}\dots\ket{\alpha_m}$, where $\bg{\alpha} = (\alpha_1,\dots,\alpha_m) \in \mathbb{C}^m$. A state is called \emph{classical} if it can be written as a probabilistic mixture of coherent states:
\begin{equation} \label{eqn:classical}
	\rho = \int \dd^{2m} \bg{\alpha}\, P(\bg{\alpha}) \proj{\bg{\alpha}}, \quad P(\bg{\alpha}) \geq 0.
\end{equation}
Due to the SSR employed here, we restrict to number-diagonal (ND) classical states -- i.e, those satisfying $[\rho,\hat{N}]=0$.

The operationally motivated free operations for non-classicality, presented in Ref.~\cite{Yadin2018Operational}, are very close to particle-separable operations. The only differences are that (i) rather than only the vacuum, any classical state may be prepared for free in a new mode and (ii) non-destructive measurements of total particle number can create non-classicality. Moreover, there is an entirely analogous protocol activating non-classicality into mode entanglement \cite{Kim2002Entanglement,Wang2002Theorem,Asboth2005Computable} (which in fact extends to more general notions of non-classicality \cite{Killoran2016Converting}). Whereas PE can be activated under particle-separable operations into SSR-entanglement, nonclassicality activates into entanglement accessible without local SSR constraints -- equivalently, entanglement which can be accessed when a shared phase reference is available.

This observation immediately implies a relation between the free states of the two resource theories: \emph{all ND classical states are particle-separable.} This follows from the fact that a classical state is always activated onto a separable state, which is always also SSR-separable, implying via Theorem \ref{thm:activation_faithful} that the input is particle-separable. In fact, this can be shown by a more direct argument, with details in Appendix \ref{appen:non-class}. Essentially, any multi-mode coherent state $\ket{\bg{\alpha}}$ can be regarded as a single-mode state -- for any choice of mode decomposition, there is always a passive linear unitary $U$ such that $U \ket{\bg{\alpha}} = \ket{\bar{\alpha}} \ket{0\dots 0}$, where $\abs{\bar{\alpha}}^2 = \sum_{i=1}^m \abs{\alpha_i}^2$. So any classical state is a probabilistic mixture of terms in which all particles occupy the same mode.

Evidently, ND classical states form a strict subset of particle-separable states. Consequently, we may say that nonclassicality is lower-bounded by PE in the sense that, for any distance measure of nonclassicality $M_\text{NC}^D$ constructed in the manner of (\ref{eqn:distance_measure}), the inequality $M_\text{NC}^D \geq M_\text{PE}^D$ holds.

What distinguishes the two sets of free states? As noted earlier, a striking property of PE is that multiple copies of a free state $\rho$ do not in general jointly form a free state. Viewed through the activation protocol, this is equivalent to saying that two copies of an SSR-separable state may be SSR-entangled. This is possible because of the way the SSR behaves for multiple copies of a system \cite{vaccaro2003entanglement,schuch2004nonlocal}. If $A$ and $B$ share two pairs of entangled systems, $(A_1,B_1)$ and $(A_2,B_2)$, then the particle number local to $A$ is $\hat{N}_A = \hat{N}_{A_1} + \hat{N}_{A_2}$ and similarly for $B$. The local SSR is applied by $\Phi_A \ox \Phi_B \neq \Phi_{A_1} \ox \Phi_{A_2} \ox \Phi_{B_1} \ox \Phi_{B_2}$. The lack of factorisation is due to degeneracy in the eigenvalues of $\hat{N}_A,\hat{N}_B$. For example, $({\ket{0}}_A {\ket{1}}_B + {\ket{1}}_A {\ket{0}}_B)/\sqrt{2}$ is entangled but SSR-separable; the two copy state
\begin{align}
	& \frac{1}{2}\left({\ket{0}}_A {\ket{1}}_B + {\ket{1}}_A {\ket{0}}_B\right)^{\ox 2} = \nonumber \\
	&\frac{1}{2}\left({\ket{00}}_A {\ket{11}}_B + {\ket{01}}_A {\ket{10}}_B + {\ket{10}}_A {\ket{01}}_B + {\ket{11}}_A {\ket{00}}_B\right)
\end{align}
is SSR-entangled thanks to correlations in the block $N_A=N_B=1$. This phenomenon is closely related to \emph{work-locking} in quantum thermodynamics, whereby coherence in one copy of a state is useless for work extraction but becomes usable in two copies \cite{Lostaglio2015Description}.

A tensor product of two classical states is always classical, hence multiple copies of an ND classical state always have zero PE. Are these the only states with this property? We first consider \emph{number-bounded} states: those for which the expansion $\sum_N p_N \rho^{(N)}$ terminates at a finite maximum. In this case, the resource content of two copies is sufficient to distinguish the classical subset of particle-separable states (note that all classical states apart from the vacuum are necessarily unbounded in number):
\begin{thm} \label{thm:two_copy_bounded}
	Two copies $\rho^{\ox 2}$ of a number-bounded state $\rho$ are particle-separable if and only if $\rho$ is the vacuum.
\end{thm}
(See the proof in Appendix \ref{appen:non-class}.) In the general unbounded case, let us first take pseudo-pure states, by which we mean those obtained by applying the SSR to a pure state: $\rho = \Phi(\proj{\psi})$. It is known that in the limit $k \to \infty$ of many copies ${\ket{\psi}}^{\ox k}$ of a pure entangled state, the SSR is effectively lifted in that the full entanglement entropy is distillable \cite{schuch2004nonlocal}. One may then argue from the activation protocol as follows: a non-classical state at the input results in entanglement at the output; many copies of this state must therefore result in an SSR-entangled state. Hence any non-classical pseudo-pure state must fail to be particle-separable with sufficiently many copies. An even stronger statement is in fact possible:
\begin{thm} \label{thm:two_copy_pure}
	Two copies $\Phi(\proj{\psi})^{\ox 2}$ of a pseudo-pure state are particle-separable if and only if $\ket{\psi}$ is classical.
\end{thm}
Therefore we see that non-classicality of any pseudo-pure state, even if particle-separable, can always be unlocked into non-zero PE by taking only two copies.

{ Finally, we prove the strongest possible connection between particle-separable and classical states, which concerns the case of arbitrarily many copies. The only assumption here is of a finite mean particle number (and, as usual, $\rho = \Phi(\rho)$).
\begin{thm} \label{thm:many_copy_bound}
    Let $\rho$ have finite mean particle number, $\Tr[\rho \hat{N}] < \infty$, and suppose that $\rho^{\ox k}$ is particle-separable for some $k$. Then the trace-distance non-classicality of $\rho$ is bounded by
    \begin{equation}
        M^{\Tr}_\mathrm{NC}(\rho) \leq \frac{1}{k}.
    \end{equation}
    Consequently, $\rho^{\ox k}$ is particle-separable for all $k$ if and only if $\rho$ is classical.
\end{thm}
The importance of this result is the realisation that every (finite mean number) non-classical state has the potential to contain particle entanglement, and thus all of the associated resource value,  once sufficiently many copies are taken. The proof (in Appendix~\ref{appen:non-class}) follows from a novel de Finetti-type theorem, which may be of independent interest.}

%%%%%%%%%%%%%%%%%%%%%%%%%%%%%%%%%%%%%%%%%%%%%%%%%%%%%%%%%%%%%%%%%%%%%%%%%%%%%%%%%%%%%%%%%%%%%%%%%%%%%%%%%%
\section{Discussion}
We have shown that entanglement between identical particles, despite its seemingly fictitious nature, is  described by a consistent resource theory whose free operations are implementable in a wide range of physical systems. Far from just an abstract quantity, this particle entanglement { can be quantified by virtue of the advantage it yields for quantum metrology, and} can be activated, via the same types of free operations, into directly accessible mode entanglement. This occurs in a setting where phase references are not easily shared between separated parties, enforcing a local SSR.

{While we have found the most general form that such an activation may take, some important questions remain open. Theorem~\ref{thm:measure_from_e_ssr_main} expresses the maximum activated SSR-entanglement from a given state as a measure of PE -- however, because of our construction we can raise the following question: What is the optimal operation to activate this entanglement? This may depend on the measure being employed, but it is plausible that such an optimal operation should be unitary; Lemma 5 in Appendix \ref{app:activating} proves a simplification from the full space of passive linear unitaries down to only one real parameter per mode, making the optimisation feasible.}
%Didnt bring up C-NOT as optimal because this isnt really an activation.

Our formulation reveals PE as fundamentally connected not only to entanglement under SSRs, but also to continuous variable non-classicality. In particular, we have shown that SSR-compliant classical states possess no PE. Consequently, PE is a stronger (rarer) resource than non-classicality. Nevertheless, by utilising multiple copies of a state, one may unlock its non-classicality into PE. { This unlocking is possible with two copies of any pure non-classical state; in general, non-classicality always results in PE after taking sufficiently many copies. Hence, in a sense, non-classicality emerges as a many-copy limit of PE. It is worth exploring other quantitative ways in which this limit may manifest itself.}

{
It is also worth noting some similarity with other resource theories. For instance, the structure of particle-separable operations bears some resemblance to ``strictly incoherent operations", a set of free operations for quantum coherence~\cite{Yadin2016Quantum}. Without measurements, particle-separable operations coincide with the zero-temperature limit of a recent treatment of continuous-variable thermodynamics~\cite{Narasimhachar2019Thermodynamic} (see also the related approach~\cite{Serafini2020Gaussian}). One could therefore explore thermodynamical consequences of PE in future work.
}

Finally, we would like to motivate the wider theoretical and experimental applicability of our framework for PE. In addition to describing the { metrological power and the} activation of entanglement from a BEC, the framework applies to any system of identical bosons, opening up the possibility of investigating PE beyond BECs and optics, to other condensed matter systems in which entanglement is of interest, such as superfluid Helium \cite{herdman2017entanglement}.

{A study of PE in fermionic systems could also be pursued, as this would have additional relevance for condensed matter. However, there are significant differences with the bosonic case. For instance, in the fermionic counterpart of the resource theory reported here, the free states, being both antisymmetric and particle-separable, would be just the single-particle and vacuum states.

It is hoped that the results presented here will stimulate further theoretical and experimental studies, across the communities of quantum information, quantum optics and condensed matter, in order to gain valuable insight into genuinely quantum properties of identical particles and their technological applications.}

%%TC:ignore
%%%%%%%%%%%%%%%%%%%%%%%%%%%%%%%%%%%%%%%%%%%%%%%%%%%%%%%%%%%%%%%%%%%%%%%%%%%%%%%%%%%%%%%%%%%%%%%%%%%%%%%%%%
\begin{acknowledgments}
We are grateful to Ludovico Lami for simplifying the proof of Theorem~\ref{thm:free_projectors} and Vlatko Vedral, Chiara Marletto, Gabriel Landi, and Richard Howl for helpful discussions. We also thank Alessia Castellini, Rosario Lo Franco, Giuseppe Compagno and the Physical Institute for Theoretical Hierarchy (PITH) for encouraging an investigation into this topic. We acknowledge financial support from the European Research Council (ERC) under the Starting Grant GQCOP (Grant No.~637352) and the EPSRC (Grant No.~EP/N50970X/1). M.F., T.Z. and P.T. acknowledge support of the Swiss National Science Foundation. M.F. acknowledges support from The National Natural Science Foundation of China (Grants No.~11622428 and No.~61675007).
\end{acknowledgments}

\appendix
 \section{{FORM OF FREE STATES}}\label{app:free_states}
Here we show that every particle-separable state of $N$ particles is of the first-quantised form
\begin{equation}
\rho^{\fq} = \sum_i \lambda_i {\proj{\psi_i}}^{\ox N}, \; \lambda_i \geq 0.
\end{equation}

By assumption, $\rho^{\fq}$ is separable, so we can write $\rho^{\fq} = \sum_i \lambda_i \bigotimes_{k=1}^N \proj{\psi_i^k}$. Since
\begin{equation}
\bigotimes_{k=1}^N \ket{\psi_i^k} \in \mathrm{supp}\, \rho^{\fq} \subseteq \mc{H}_N,
\end{equation}
each term $\bigotimes_{k=1}^N \ket{\psi_i^k}$ is in the symmetric subspace. It follows from this symmetry that all $\ket{\psi_i^k}$ are the same for a given $i$.

%%%%%%%%%%%%%%%%%%%%%%%%%%%%%%%%%%%%%%%%%%%%%%%%%%%%%%%%%%%%%%%%%%%%%%%%%%%%%%%%%%%%%%%%%%%%%%%%%%%%%%%%%%
\section{APPENDING FREE STATES}\label{app:appending}

\begin{thm}\label{thm:appending_vaccuum}
	The operation $\mc{E}(\rho) = \rho \ox \sigma$, which appends a fixed state $\sigma$ in a new set of $m$ modes, preserves the set of free states if and only if $\sigma = \proj{0}$.
	\begin{proof}
		It is sufficient to let $\rho$ be the simplest free state, a single particle in a single mode: $\rho=\proj{1}$. $\sigma=\sum_N p_N \sigma^{(N)}$ is arbitrary and may have unbounded particle number. Then
		\begin{align}
		\rho\ox\sigma=\sum_N p_N \proj{1} \ox \sigma^{(N)}.
		\end{align}
		The $(N+1)$-particle component of this state is $\proj{1} \ox \sigma^{(N)}$. In order to particle-separable, it must be possible to express as
		\begin{equation}
		\proj{1} \ox \sigma^{(N)} = \sum_i \lambda_i \, U_i {\proj{N+1,0,0\dots}} U_i^\dagger,
		\end{equation}
		in terms of some set of $m+1$ modes, with $\lambda_i \geq 0$ and the $U_i$ being free unitaries. The left-hand side has exactly one particle in the first mode and $N$ in the remainder, so the same must be true of every term on the right-hand side. So for each $i$, $U_i \ket{N+1,0,\dots} = \ket{1}\ket{\psi_i}$, which is impossible unless $N=0$. To see this, note that we can write
		\begin{equation}
		U_i \ket{N+1,0\dots} \propto (a_1^\dagger + b_i^\dagger)^{N+1} \ket{0},
		\end{equation}
		where $b_i$ is some linear combination of annihilation operators on the rightmost $N$ modes. Expanding the bracket $(a_1^\dagger + b_i^\dagger)^{N+1}$, we can never have a single term linear in $a_1^\dagger$ unless $N=0$.
		
		Therefore $p_N = 0$ for $N \geq 0$, so $\sigma = \proj{0}$. Conversely, it is trivially seen that appending vacuum modes always preserves the set of free states.
	\end{proof}
\end{thm}	
%%%%%%%%%%%%%%%%%%%%%%%%%%%%%%%%%%%%%%%%%%%%%%%%%%%%%%%%%%%%%%%%%%%%%%%%%%%%%%%%%%%%%%%%%%%%%
\section{FREE UNITARIES}\label{app:unitaries}
In the following section, we work with states of $N$ particles and always in the first-quantised picture, so we drop the additional notation for convenience.

\begin{thm}
	A unitary $U$ on $\mc{H}_N$ maps free states into free states if and only if $U = u^{\ox N}$.
	\begin{proof}
		Note that we only specify the restriction of $U$ to $\mc{H}_N$ rather than the ``full" Hilbert space $\mc{H}_1^{\ox N}$. For example, permutations between particles are not of the given form but have trivial action on the symmetric subspace.
		
		By assumption, for any $\ket{\Psi} = {\ket{\psi}}^{\ox N}$, we have $U \ket{\Psi} = \ket{\Phi(\Psi)} := {\ket{\phi(\psi)}}^{\ox N}$. Taking an inner product for two arbitrary $\psi,\psi'$:
		\begin{equation}
		\braket{\Psi'}{\Psi} = \braket{\Phi(\Psi')}{\Phi(\Psi)} \Rightarrow {\braket{\psi'}{\psi}}^{N} = {\braket{\phi(\psi')}{\phi(\psi)}}^N.
		\end{equation}
		The $N$th root of this gives
		\begin{eqnarray}
		&&\braket{\phi(\psi')}{\phi(\psi)} = \braket{\psi'}{\psi} e^{2\pi i n(\psi,\psi') /N}, \\ && \nonumber \qquad n(\psi,\psi') \in \{0,1,\dots,N-1\}.
		\end{eqnarray}
		Both sides of this equation must be continuous in $\psi,\psi'$. But $n(\psi,\psi')$ is a continuous integer-valued function, so must be constant. In particular, $n(\psi, \psi) = 0$, so we conclude that $n \equiv 0$.
		
		By Wigner's theorem \cite{wignerstuff}, any transformation of states that preserves the inner product must be unitary. Therefore there exists unitary $u$ such that $\ket{\phi(\psi)} = u \ket{\psi} \, \forall \psi$, which proves the result.
	\end{proof}
\end{thm}
%%%%%%%%%%%%%%%%%%%%%%%%%%%%%%%%%%%%%%%%%%%%%%%%%%%%%%%%%%%%%%%%%%%%%%%%%%%%%%%%%%%%%%%%%%%%%%%%%%%%%%%%%%%%%%%%%%%%%%%%%%
\section{ FREE MEASUREMENTS}\label{app:measurements}
As in Appendix \ref{app:unitaries}, we temporarily drop the first-quantised notation. As a first step in the investigation of non-destructive measurements, we need the following Lemma:
\begin{lem}\label{lem:single_projector}
	Let $\Pi$ be a projector with support on the symmetric subspace of $N$ particles, i.e.\ $\Pi = P_N \Pi P_N$, where $P_N$ projects onto $\mc{H}_N$. Then $\Pi$ is non-entangling if and only if there exists a projector $\pi$ on $\mc{H}_1$ such that
	\begin{equation} \label{eqn:product_projectorb}
	\Pi = P_N \pi^{\ox N} P_N .
	\end{equation}
	\begin{proof}
		It is immediate that any $\Pi$ of the form (\ref{eqn:product_projectorb}) preserves symmetric product states; so we need only prove the converse. We start from the observation that for any $\ket{\psi} \in \mc{H}_1$, there is a (normalised) $\ket{\phi} \in \mc{H}_1$ such that $\Pi {\ket{\psi}}^{\ox N} = c {\ket{\phi}}^{\ox N}$, where either $c = 0$ or else $c\neq 0$ and ${\ket{\phi}}^{\ox N} \in \supp \Pi$. If $c=0\; \forall \ket{\psi}$, then $\Pi=0$ since states of the form ${\ket{\psi}}^{\ox N}$ span $\mc{H}_N$ \cite{harrow2013church}. Otherwise, there must exist some $\ket{0}$ such that ${\ket{0}}^{\ox N} \in \supp \Pi$.
		
		If $\rank \Pi = 1$, then $\Pi = {\proj{0}}^{\ox N}$ and we are done. If $\rank \Pi >1$, then consider any $\ket{\psi}$ orthogonal to $\ket{0}$. Again, we must have $\Pi {\ket{\psi}}^{\ox N} = c {\ket{\phi}}^{\ox N}$. Note that
		\begin{align}
		c {\braket{0}{\phi}}^N & = c {\bra{0}}^{\ox N} {\ket{\phi}}^{\ox N} \nonumber \\
		& = {\bra{0}}^{\ox N} \left( \Pi {\ket{\psi}}^{\ox N} \right) \nonumber \\
		& = {\bra{0}}^{\ox N} {\ket{\psi}}^{\ox N} = 0,
		\end{align}
		having used $\Pi {\ket{0}}^{\ox N} = {\ket{0}}^{\ox N}$. So either $c = 0$, or else $c\neq 0$ and $\ket{\phi}$ is orthogonal to $\ket{0}$. Considering all $\ket{\psi}$ orthogonal to $\ket{0}$, it follows that either $\Pi {\ket{\psi}}^{\ox N} = 0$ for all such $\ket{\psi}$, or else there exists $\ket{1}$ orthogonal to $\ket{0}$, with ${\ket{1}}^{\ox N} \in \supp \Pi$.
		
		Continuing this procedure, we are able to construct a complete basis $\{ \ket{k} \}$ of $\mc{H}_1$ such that
		\begin{equation}
		{\ket{k}}^{\ox N} \in
		\begin{cases}
		\supp \Pi,	& 0 \leq k \leq r-1 \\
		\ker \Pi,	& r \leq k \leq d-1
		\end{cases}
		\end{equation}
		for some $r$.
		
		Now take an arbitrary $\ket{\psi} \in \mc{H}_1$, written in terms of the chosen basis as $\ket{\psi} = \sum_{k=0}^{d-1} \psi_k \ket{k}$. Given the properties of this basis, it follows that
		\begin{equation}
		{\bra{k}}^{\ox N} \Pi {\ket{\psi}}^{\ox N} =
		\begin{cases}
		{\bra{k}}^{\ox N}{\ket{\psi}}^{\ox N} = \psi_k^N,	& 0 \leq k \leq r-1 \\
		0,	& r \leq k \leq d-1.
		\end{cases}
		\end{equation}
		But since $\Pi$ preserves product states, $\Pi {\ket{\psi}}^{\ox N} = {\ket{\phi}}^{\ox N}$ (where $\ket{\phi}$ need not be normalised). Expressing $\ket{\phi} = \sum_{k=0}^{d-1} \phi_k \ket{k}$, ${\bra{k}}^{\ox N} {\ket{\phi}}^{\ox N} = \phi_k^N$, thus
		\begin{equation}
		\phi_k =
		\begin{cases}
		\psi_k e^{2\pi i n_k/N}	& 0 \leq k \leq r-1 \\
		0,	& r \leq k \leq d-1,
		\end{cases}
		\end{equation}
		where $n_k \in \{0,\dots,N-1\}$. In principle, $n_k$ may be a function of $\ket{\psi}$; however, the continuity of the mapping under $\Pi$ ensures that $n_k$ is continuous and hence constant. Furthermore, since ${\ket{\phi}}^{\ox N}$ is invariant under this mapping, we must have $n_k \equiv 0$, so that $\phi_k = \psi_k \; \forall k \leq r-1$.
		
		The action of $\Pi$ on an arbitrary product ${\ket{\psi}}^{\ox N}$ is therefore identical to the action of $\pi^{\ox N}$, where
		\begin{equation}
		\pi := \sum_{k=0}^{r-1} \proj{k}.
		\end{equation}
		Again, since such product states span $\mc{H}_N$, this gives (\ref{eqn:product_projectorb}).
	\end{proof}
\end{lem}

\begin{thm}\label{thm:free_projectors}
	Let $\{\Pi_i\}_{i=1}^k$ be a set of non-zero orthogonal projectors onto subspaces of $\mc{H}_N$ (where $N > 1$) such that $\sum_{i=1}^k \Pi_i = P_N$ and each $\Pi_i$ preserves the set of particle-separable states. Then $k=1$ and
	\begin{align}
	\Pi_1 = P_N.
	\end{align}
	\begin{proof}
		From Lemma~\ref{lem:single_projector}, there exist projectors $\pi_i$ such that $\Pi_i = P_N \pi_i^{\ox N} P_N \, \forall i$. It follows from this that the orthogonality relation $\Pi_i \Pi_j = \delta_{i,j} \Pi_i$ implies $\pi_i \pi_j = \delta_{i,j} \pi_i$. Hence there exist orthogonal $\ket{\psi_i}$ such that $\ket{\psi_i} \in \supp \pi_i$. From these, we construct $\ket{\psi} := \frac{1}{\sqrt{k}} \sum_{i=1}^k \ket{\psi_i}$. The action of $\Pi_i$ on ${\ket{\psi}}^{\ox N}$ is
		\begin{equation}
		\Pi_i {\ket{\psi}}^{\ox N} = (\pi \ket{\psi})^{\ox N},
		\end{equation}
		from which the completeness relation gives
		\begin{equation}
		1 = \sum_{i=1}^k {\bra{\psi}}^{\ox N} \Pi_i {\ket{\psi}}^{\ox N} = \sum_{i=1}^k {\bra{\psi}\pi_i \ket{\psi}}^N.
		\end{equation}
		Using the form of $\ket{\psi}$, the right-hand side evaluates to
		\begin{equation}
		\sum_{i=1}^k {\bra{\psi} \pi_i \ket{\psi}}^N = \sum_{i=1}^k \left( \frac{1}{k} \right)^N = \frac{1}{k^{N-1}}.
		\end{equation}
		Hence there is a contradiction unless $k=1$, which forces the single projector to be $\Pi_1 = P_N$.
	\end{proof}
\end{thm}

Theorem~\ref{thm:free_projectors} says that any non-destructive free projective measurement in the $N$-particle subspace must be trivial. Extending this to measurements over the whole Fock space, respecting the SSR, shows that only a measurement of the number observable $\hat{N}$ is permissible.

\begin{thm}\label{thm:destructive_measurement}
	Any destructive measurement respecting the SSR preserves the set of particle-separable states $\mc{S}$.
	\begin{proof}
		It is sufficient to prove this for a single projector. Let the measurement be performed on $m_B$ modes of an $(m_A+m_B)$-mode system, having the action
		\begin{equation}
		\rho_{AB} \to \sigma_A = \Tr_B \left[ (\id_A \ox \Pi_B) \rho_{AB} \right],
		\end{equation}
		where $\Pi_B$ is a projector such that $[\Pi_B,\hat{N}_B]=0$. Any particle-separable pure state has the form $\ket{\psi} \propto (c^\dagger)^N \ket{0}$, where $c$ is a single-particle annihilation operator. Choosing some orthogonal mode set $\{a_i\}$, where $i = 1,\dots,m_A$ for the unmeasured modes and $i=m_A+1,\dots,m_A+m_B$ for the measured modes, we can write $c = a + b$, where $a$ and $b$ are linear combinations of the unmeasured and measured $a_i$, respectively. Thus we can effectively treat $\ket{\psi}$ as a two-mode state:
		\begin{align}
		\ket{\psi} & = \left(a^\dagger + b^\dagger \right)^N {\ket{0}}_A{\ket{0}}_B \nonumber \\
		& = \sum_{N_A} r_{N_A} {\ket{N_A}}_A {\ket{N-N_A}}_B,
		\end{align}
		where the $r_{N_A}$ are coefficients.
		
		Then the post-measurement (unnormalised) state is
		\begin{align}
		\sigma_A  =& \Tr_B \Big{[} \sum_{N_A,N_A'} r_{N_A} r_{N_A'}^* (\id_A \ox \Pi_B) {\ketbra{N_A}{N_A'}}_A\nonumber\\
		 &\ox {\ketbra{N-N_A}{N-N_A'}}_B \Big{]} \nonumber \\
		=& \sum_{N_A,N_A'} r_{N_A} r_{N_A'}^* {\bra{N-N_A'}}_B \Pi_B {\ket{N-N_A}}_B {\ketbra{N_A}{N_A'}}_A \nonumber \\
		=& \sum_{N_A,N_A'} r_{N_A} r_{N_A'}^* s_{N_A} \, \delta_{N_A,N_A'} {\ketbra{N_A}{N_A'}}_A \nonumber \\
		=& \sum_{N_A} \abs{r_{N_A}}^2 s_{N_A} \, {\ketbra{N_A}{N_A}}_A.
		\end{align}
		where we have used the fact that $\Pi_B$ is diagonal in particle number, $[\Pi_B,\hat{N}_B]=0$, to give ${\bra{M}}_B \Pi_B {\ket{N}}_B = s_N \delta_{N,M}$. Hence $\sigma_A \in \mc{S}$; the extension to mixed initial states $\rho_A$ follows by linearity.
	\end{proof}
\end{thm}

%%%%%%%%%%%%%%%%%%%%%%%%%%%%%%%%%%%%%%%%%%%%%%%%%%%%%%%%%%%%%%%%%%%%%%%%%%%%%%%%%%%%%%%%%%%%%%%%%%%%%%%%%%%%%%%%%%%%%%%%%%
\section{MEASURES OF PE}\label{app:measures}
The following results are used to show that if $D$ satisfies a few straightforward properties, then the resulting measure of PE can be expressed as an average over different particle numbers. We write this in a more abstract form which shows a generalisation to arbitrary resource theories with a block-diagonal structure.
\begin{lem} \label{lem:distances}
	Suppose a distance measure $D$ satisfies
	\begin{enumerate}
		\item (contractivity) $D(\mc{E}(\rho),\mc{E}(\sigma)) \leq D(\rho,\sigma)$ under any channel $\mc{E}$;
		\item (joint convexity) $D(\sum_i p_i \rho_i, \sum_i p_i \sigma_i) \leq \sum_i p_i D(\rho_i,\sigma_i)$ for any sets of states $\rho_i,\sigma_i$ and probabilities $p_i$;
		\item (direct sum concavity) $D( \bigoplus_i p_i \rho_i, \bigoplus_i q_i \sigma_i) \geq \sum_i p_i D(\rho_i, \sigma_i)$.
	\end{enumerate}
	Then it also satisfies
	\begin{enumerate}
		\item[a.] (direct sum linearity) $D(\bigoplus_i p_i \rho_i, \bigoplus_i p_i \sigma_i) = \sum_i p_i D(\rho_i,\sigma_i)$;
		\item[b.] (ensemble contractivity) $\sum_i p_i D(\rho_i,\sigma_i) \leq D(\rho,\sigma)$, where $\{\mc{E}_i\}$ is any quantum instrument, and $\mc{E}_i(\rho) = p_i \rho_i,\, \mc{E}_i(\sigma) = q_i \sigma_i$.
	\end{enumerate}
	\begin{proof}
		To show (a):
		\begin{align*}
		\sum_i p_i D(\rho_i, \sigma_i)  & \underset{(3)}{\leq} D\left(\bigoplus_i p_i \rho_i, \bigoplus_i p_i \sigma_i\right) \\
		& = D \left( \sum_i p_i \rho_i \ox \proj{i}, \sum_i p_i \sigma_i \ox \proj{i} \right) \\
		& \underset{(2)}{\leq} \sum_i p_i D \left( \rho_i \ox \proj{i}, \sigma_i \ox \proj{i} \right) \\
		& \underset{(1)}{=} \sum_i p_i D(\rho_i, \sigma_i),
		\end{align*}
		where, in the last line, we have used the fact that adding and removing an uncorrelated system are both reversible channels which must therefore leave $D$ unchanged. The left- and right-hand sides are equal, thus the initial inequality must actually be an equality.
		
		To show (b), we construct from the instrument a channel $\mc{E}(\rho) = \sum_i \mc{E}_i(\rho) \ox \proj{i}$, so that
		\begin{align*}
		\sum_i p_i D(\rho_i,\sigma_i) & \underset{(3)}{\leq}  D \left( \bigoplus_i p_i \rho_i, \bigoplus_i q_i \sigma_i \right) \\
		& = D \left( \sum_i p_i \rho_i \ox \proj{i} , \sum_i q_i \sigma_i \ox \proj{i} \right) \\
		& = D \left( \mc{E}(\rho), \mc{E}(\sigma) \right) \\
		& \underset{(1)}{\leq} D(\rho,\sigma).
		\end{align*}
	\end{proof}
\end{lem}

From this, we obtain:
\begin{thm} \label{thm:block_measures}
	Suppose that $D$ satisfies properties (1,2,3) listed in Lemma \ref{lem:distances}. Let $F$ be any convex set of states, and define
	\begin{equation}
	M^D(\rho) := \min_{\sigma \in F} D(\rho,\sigma).
	\end{equation}
	Then $M^D$ is an ensemble monotone under instruments $\{\mc{E}_i\}$ such that each $\mc{E}_i$ preserves the set $F$.
	
	Furthermore, if $F = \bigoplus_N F_N$, where each $F_N$ is a convex set of states, then
	\begin{equation}
	M^D \left( \bigoplus_N p_N \rho^{(N)} \right) = \sum_N p_N M^D_N( \rho^{(N)} ),
	\end{equation}
	where $M^D_N$ is defined similarly to $M^D$, but minimising over states in $F_N$.
	\begin{proof}
		For the first part, we take $\tau$ to be the closest state to $\rho$ in $F$. For any instrument $\{\mc{E}_i\}$, let $p_i \rho_i = \mc{E}_i(\rho),\, q_i \tau_i = \mc{E}_i(\tau)$. Then
		\begin{align*}
		M^D(\rho) & = D(\rho, \tau) \\
		& \underset{(b)}{\geq} \sum_i p_i D(\rho_i, \tau_i) \\
		& \geq \sum_i p_i \min_{\sigma_i \in F} D(\rho_i,\sigma_i) \\
		& = \sum_i p_i M^D(\rho_i).
		\end{align*}
		For the second part,
		\begin{align*}
		M^D& \left( \bigoplus_N p_N \rho^{(N)} \right) \nonumber\\
		&= \min_{ \{ q_N, \, \sigma^{(N)} \in F_N \} } D \left( \bigoplus_N p_N \rho^{(N)} , \bigoplus_N q_N \sigma^{(N)} \right) \\
		& \underset{(3)}{\geq} \sum_N p_N \min_{ \sigma^{(N)} \in F_N} D \left( \rho^{(N)}, \sigma^{(N)} \right) \\
		& \underset{(2)}{\geq} \min_{ \{ \sigma^{(N)} \in F_N \} } D \left( \bigoplus_N p_N \rho^{(N)}, \bigoplus_N p_N \sigma^{(N)} \right),
		\end{align*}
		which shows that the closest state can be chosen to have $q_N = p_N$. Finally, we use (a).
	\end{proof}
\end{thm}

The relative entropy $S(\rho||\sigma) := \Tr[ \rho \log \rho - \rho \log \sigma]$ satisfies all three assumptions of Lemma~\ref{lem:distances} -- in particular, (3) follows from
\begin{equation}
S\left(\bigoplus_i p_i \rho_i || \bigoplus_i q_i \sigma_i\right) = \sum_i p_i S(\rho_i || \sigma_i) + H( \{p_i\}|| \{q_i\} ),
\end{equation}
where the last term is the classical relative entropy (or Kullback-Leibler divergence). Hence the relative entropy measure of PE is
\begin{equation}
M^{RE}_{\text{PE}}(\rho) = \sum_N p_N M^{RE}_{\text{PE}}(\rho^{(N)}).
\end{equation}
The same property also holds for distances defined by Schatten $p$-norms, $D_p(\rho,\sigma) = \| \rho - \sigma \|_p$ \cite{chitambar2019quantum}.

%%%%%%%%%%%%%%%%%%%%%%%%%%%%%%%%%%%%%%%%%%%%%%%%%%%%%%%%%%%%%%%%%%%%%%%%
%%%%%%%%%%%%%%%%%%%%%%%%%%%%%%%%%%%%%%%%%%%%%%%%%%%%%%%%%%%%%%%%%%%%%%%%%%%%%%%%%%%%%%%%%%%%%%%%%%%%%%%%%%%%%%%%%%%%%%%%%

\section{MONOTONICITY OF METROLOGICAL MEASURE} \label{app:qfi}
The proof of monotonicity of $\mqfi$ makes use of the following Lemma (which is to our knowledge novel):
\begin{lem} \label{lem:qfi_projector}
	Let $\Pi$ be a projector such that $\Pi \rho = \rho$. Then
	\begin{equation}
		\mc{F}(\rho,H) = \mc{F}(\rho, \Pi H \Pi) + 4V(\rho,H) - 4V(\rho, \Pi H \Pi).
	\end{equation}
	\begin{proof}
		Given the spectral decomposition $\rho = \sum_{i=0}^{d-1} \lambda_i \proj{i}$, we have $\lambda_i \Pi \ket{i} = \Pi \rho \ket{i} = \lambda_i \ket{i}$, so $\Pi \ket{i} = \ket{i}\; \forall \ket{i} \in \mathrm{supp}\, \rho$. Therefore we can write $\Pi = \sum_{i<r} \proj{i}$, such that $\lambda_j = 0 \; \forall j \geq r$, where $r = \mathrm{rank}\, \Pi \geq \mathrm{rank}\, \rho$. It follows that
		\begin{align}
			\mc{F}(\rho,H) & = 2 \sum_{i,j} \frac{(\lambda_i-\lambda_j)^2}{\lambda_i+\lambda_j} \abs{\bra{i}H\ket{j}}^2 \nonumber \\
				& = 2 \sum_{i,j<r} \frac{(\lambda_i-\lambda_j)^2}{\lambda_i+\lambda_j} \abs{\bra{i}H\ket{j}}^2 \nonumber \\
					& \qquad + 4 \sum_{i<r, j\geq r} \frac{(\lambda_i-0)^2}{\lambda_i+0} \abs{\bra{i}H\ket{j}}^2 \nonumber \\
				& = 2 \sum_{i,j<r} \frac{(\lambda_i-\lambda_j)^2}{\lambda_i+\lambda_j} \abs{\bra{i}\Pi H \Pi\ket{j}}^2 \nonumber \\
				& \qquad + 4 \sum_{i<r, j\geq r} \frac{(\lambda_i-0)^2}{\lambda_i+0} \abs{\bra{i}H\ket{j}}^2 \nonumber \\
				& = \mc{F}(\rho, \Pi H \Pi) + 4 \sum_{i<r, j\geq r} \lambda_i \bra{i}H\ket{j}\bra{j}H\ket{i} \nonumber \\
				& = \mc{F}(\rho, \Pi H \Pi ) + 4 \sum_{i<r} \lambda_i \bra{i}H \left(\sum_{j\geq r} \proj{j} \right)H\ket{i} \nonumber \\
				& = \mc{F}(\rho, \Pi H \Pi) + 4 \sum_{i<r} \lambda_i \bra{i}H(I - \Pi)H\ket{i} \nonumber \\
				& = \mc{F}(\rho, \Pi H \Pi) + 4 \Tr( \rho H^2) - 4\Tr(\rho H \Pi H) \nonumber \\
				& = \mc{F}(\rho, \Pi H \Pi) + 4 \Tr( \rho H^2) - 4\Tr(\rho [\Pi H \Pi]^2) \nonumber \\
				& = \mc{F}(\rho, \Pi H \Pi) + 4 V(\rho,H) - 4V(\rho, \Pi H \Pi),
		\end{align}
		where the last line uses $\Tr(\rho \Pi H \Pi) = \Tr(\rho H)$.
	\end{proof}
\end{lem}

\begin{manualtheorem}{\ref{thm:qfi_monotone}}[{\bf main text}]
    $\mqfi$ is convex and satisfies $\mqfi(\rho) = 0 \; \forall \rho \in \mc{S}$.
    Moreover, let $\mc{E}_{S \to SM} \in \mc{O}$ contain a single measurement round, such that no conditional operations are performed after the measurement. We may write $\mc{E}_{S\to SM}(\rho_S) = \sum_m \mc{E}^m(\rho_S) \ox {\proj{m}}_M$, where $\mc{E}^m$ is the operation applied to $\rho_S$ conditioned on outcome $m$. Then
    \begin{equation}
        \mqfi(\rho_S) \geq \mqfi(\mc{E}_{S \to SM}[\rho]).
    \end{equation}

    \begin{proof}
        Convexity of $\mqfi$ follows from convexity of both the QFI and the function $\pospart{\cdot}$, and concavity of the variance:
        \begin{align}
        	\mqfi(p \rho + (1-p) \sigma) & \leq \max_h \left[p \mc{F}(\rho,H) + (1-p)\mc{F}(\sigma,H)\right. \nonumber \\
            	& \quad \left. - 4pV(\rho,h) -4(1-p)V(\sigma,h)\right]^+ \nonumber \\
        		& \leq \max_h p \pospart{\mc{F}(\rho,H) - 4V(\rho,h)} \nonumber \\
        		& \quad + (1-p) \pospart{\mc{F}(\sigma,H)-4V(\sigma,h)} \nonumber \\
        		& \leq p \mqfi(\rho) + (1-p) \mqfi(\sigma).
        \end{align}

        We break the proof of monotonicity into the three stages of a particle-separable operation without feed-forward: i) appending modes in the vacuum state; ii) performing a global passive linear unitary; iii) destructively measuring a set of modes.

        \emph{i) Appending modes in the vacuum state:} We append to the system modes $S$ a set of vacuum ancilla modes $A$. Our aim is to show that
        	\begin{equation} \label{eqn:qfi_vacuum_inv}
		        \mqfi(\rho_S \ox {\proj{0}}_A) = \mqfi(\rho_S).
	        \end{equation}
	   The proof consists of showing that the optimal observable for the vacuum-added state always acts solely on $S$. Note that the single-particle Hilbert space of $SA$ splits into $\mc{H}_1 = \mc{H}_{1,S} \oplus \mc{H}_{1,A}$; we denote the projectors onto these subspaces by $\Pi_S,\Pi_A$ respectively. Thus any $h$ can be decomposed into the terms
		\begin{equation}
			h = \Pi_S h \Pi_S + \Pi_A h \Pi_A + \Pi_S h \Pi_A + \Pi_A h \Pi_S =: h' + g' + f + f^\dagger.
		\end{equation}
		Each term gives rise to its own second-quantised observable exactly as for $H$, i.e.\ ${H'}^{(N)\fq} = \sum_{i=1}^N h'_i/\sqrt{N}$ and so on.
		
		We apply Lemma~\ref{lem:qfi_projector} using $H$ and the projector $\Pi = I_S \ox {\proj{0}}_A$. It may be seen that in first quantisation, $\Pi^{(N)\fq} = \Pi_S^{\ox N}$, so that each particle is projected on the subspace $\mc{H}_{1,S}$. Therefore we see that $\Pi H \Pi = H'$. Thus
		\begin{align}
			\mc{F}(\rho_S \ox {\proj{0}}_A,H) & = \mc{F}(\rho_S, H') + 4V(\rho_S \ox {\proj{0}}_A, H) \nonumber \\
			    & \quad - 4V(\rho_S, H') \nonumber \\
				& = \mc{F}(\rho_S,H') + 4\Tr(\rho_S \ox {\proj{0}}_A H^2) \nonumber \\
				& \quad - 4\Tr(\rho_S {H}^2) \nonumber \\
				& = \mc{F}(\rho_S,H') \nonumber \\
				& \quad + 4\Tr(\rho_S \ox {\proj{0}}_A [\Pi H^2 \Pi - {H'}^2])
		\end{align}
		using $\Tr(\rho_S \ox {\proj{0}}_A H) = \Tr(\rho_S H')$ for the second line. Now one can also see that $\Pi H^2 \Pi = {H'}^2 + \Pi F F^\dagger \Pi$, so
		\begin{equation}
			\mc{F}(\rho_S \ox {\proj{0}}_A,H) = \mc{F}(\rho_S,H') + 4 \Tr(\rho_S \ox {\proj{0}}_A \Pi F F^\dagger \Pi ).
		\end{equation}
		From $(F F^\dagger)^{(N)\fq} = \frac{1}{N} \sum_{i,j=1}^N f_i f_j^\dagger$, it follows that
		\begin{equation}
			\Pi_S^{\ox N} (F F^\dagger)^{(N)\fq} \Pi_S^{\ox N} = \frac{1}{N} \sum_{i=1}^N f_if_i^\dagger,
		\end{equation}
		since $\Pi_S f \Pi_S = 0$ but $\Pi_S f f^\dagger \Pi_S \neq 0$ . Consequently,
		\begin{equation}
			\Tr(\rho_S \ox {\proj{0}}_A \Pi F F^\dagger \Pi) = \expect{f f^\dagger}_{\rho_S \ox {\proj{0}}_A}.
		\end{equation}
		Next we have
		\begin{align} \label{eqn:qfi_plus_vacuum}
			\mc{F}(\rho_S \ox {\proj{0}}_A, H) - 4V(\rho_S \ox {\proj{0}}_A,h_1) & \nonumber \\
			& \hspace{-15em} = \mc{F}(\rho_S,H') + 4 \expect{f f^\dagger - h^2}_{\rho_S \ox {\proj{0}}_A} + 4 \expect{h}^2_{\rho_S \ox {\proj{0}}_A} \nonumber \\
				& \hspace{-15em} = \mc{F}(\rho_S,H') - 4\expect{{h'}^2}_{\rho_S} + 4 \expect{h'}^2_{\rho_S} \nonumber \\
				& \hspace{-15em} = \mc{F}(\rho_S,H') - 4V(\rho_S,h').
		\end{align}
		Now $\|h'\| = \|\Pi_S h \Pi_S\| \leq \|h\| \|\Pi_S\| = \|h\|$. If $\|h'\|=0$, then both sides of \eqref{eqn:qfi_plus_vacuum} are zero and there is nothing left to prove; otherwise, we define $\tilde{h} := h'/ \|h'\|$, which has unit norm. Putting this into \eqref{eqn:qfi_plus_vacuum} gives
		\begin{align}
			\mc{F}(\rho_S \ox {\proj{0}}_A, H) - 4V(\rho_S \ox {\proj{0}}_A,h) & \nonumber \\
			    & \hspace{-15em} = \|h'\|^2 \left[ \mc{F}(\rho_S,\tilde{H}) - 4V(\rho_S,\tilde{h}) \right] \nonumber \\
				& \hspace{-15em} \leq \pospart{ \mc{F}(\rho_S,\tilde{H}) - 4V(\rho_S,\tilde{h})} \nonumber \\
				& \hspace{-15em} \leq \mqfi(\rho_S).
		\end{align}
		Maximising over $h$ gives $\mqfi(\rho_S \ox {\proj{0}}_A) \leq \mqfi(\rho_S)$. Conversely, it is clear that equality is obtained by taking for $\rho_S \ox {\proj{0}}_A$ the same observable that maximises the quantity for $\rho_S$. Thus we have established \eqref{eqn:qfi_vacuum_inv}.
	
	    \emph{ii) Passive linear unitaries:} $\mqfi$ is explicitly invariant under such unitaries, since these correspond to a rotation of the single-particle basis, and thus just a basis change for $h$.
	
	    \emph{iii) Destructive measurement:} We start with a state $\rho_{SA}$ on two sets of modes $S,A$, where the latter ancilla modes are to be measured with a complete POVM $\{E_m\}_M$ respecting the particle-number SSR. The measurement is represented with a quantum-classical channel taking $A$ to a classical memory $M$:
        \begin{equation}
        	\rho'_{SM} := \mc{E}_{A \to M}(\rho_{SA}) := \sum_m \Tr_A[E_{m,A} \rho_{SA}] \ox {\proj{m}}_M.
        \end{equation}
        For any given $h$ acting only on $S$, we have
        \begin{align}
        	\mqfi(\rho_{SA}) & \geq \pospart{\mc{F}(\rho_{SA},H) - 4V(\rho_{SA},h)} \nonumber \\
        		& \geq \pospart{ \mc{F}(\rho'_{SM},H) - 4V(\rho_{SA},h)}.
        \end{align}
        The second inequality follows from the property of $\mc{F}(\rho,H)$ being monotonically non-increasing under operations covariant with respect to the observable $H$~\cite{yadin2016general}. Here, covariance holds because $\mc{E}_{A\to M}$ acts on a different subsystem from $H$. Next, we see that the variance part is unchanged since the statistics of $h$ do not depend on operations performed on subsystem $A$, so
        \begin{equation}
        	\mqfi(\rho_{SA}) \geq \pospart{\mc{F}(\rho'_{SM}) - 4V(\rho'_{SM},h)}.
        \end{equation}
        Finally, maximising the right-hand side over all $h$ gives $\mqfi(\rho_{SA}) \geq \mqfi(\rho'_{SM})$
    \end{proof}
\end{manualtheorem}

\color{black}
%%%%%%%%%%%%%%%%%%%%%%%%%%%%%%%%%%%%%%%%%%%%%%%%%

\section{SSR-ENTANGLEMENT}\label{app:ssr_entanglement}
The activation protocol converts particle entanglement into entanglement that is of use to two parties $A,B$ who are limited to local \emph{covariant} operations that respect the SSR and classical communication.
\begin{defn}\cite{schuch2004quantum,schuch2004nonlocal}
	An operation between two or more parties is said to be \emph{covariant-LOCC} when it is composed of local operations respecting the local superselection rule, and classical communication.
\end{defn}
Although not spelled out explicitly by \cite{schuch2004quantum,schuch2004nonlocal}, the free states of this resource theory (in a bipartite setting; easily generalised) are the following:
\begin{defn}
	A bipartite state $\rho_{AB}$ is free in the resource theory of SSR-entanglement when it can be written in the form
	\begin{equation}
	\rho_{AB} = \sum_i p_i \rho_A^i \ox \rho_B^i
	\end{equation}
	such that each $\rho_A^i,\rho_B^i$ respects the SSR, i.e., $\Phi_S(\rho_S^i)=\rho_S^i$, $S = A,B$. Such a free state is said to be \emph{invariant-separable} (since it is invariant under local phase rotations).
\end{defn}
Of course every invariant-separable state is separable, but not vice-versa. This set of free states may be motivated as being those accessible from a given primitive state, such as the vacuum $\ket{0}\ket{0}$ under covariant-LOCC.

\begin{lem} \label{lem:ssr_separable}
	The following statements are equivalent:
	\begin{enumerate}
		\item $\rho_{AB}$ is invariant-separable.
		\item $\rho_{AB} = \sum_i p_i \psi_A^i \ox \psi_B^i$ where each $\psi_A^i,\psi_B^i$ is pure and contains a definite number of particles.
		\item $\rho_{AB}$ is separable and satisfies the local SSR constraint $(\Phi_A \ox \Phi_B)(\rho_{AB}) = \rho_{AB}$.
		\item $(\Phi_A \ox \Phi_B)(\rho_{AB}) = \rho_{AB}$ and, for each $N_A,N_B$, the local-number projected state $(P_{N_A} \ox P_{N_B}) \rho_{AB} (P_{N_A} \ox P_{N_B})$ is separable.
	\end{enumerate}
	\begin{proof}
		The equivalence of (1) and (2) is easily seen from the fact that every local-SSR-respecting state $\rho_A^i = \Phi_A(\rho_A^i)$ can be written as a mixture of pure states of definite number. (1) $\Rightarrow$ (3) is also straightforward. Conversely, suppose (3) holds, then we have $\rho_{AB} = \sum_i p_i \rho_A^i \ox \rho_B^i$ for arbitrary states $\rho_A^i,\rho_B^i$. But then the local SSR constraint implies that $\rho_{AB} = \sum_i p_i \sigma_A^i \ox \sigma_B^i$, where $\sigma_S^i = \Phi_S(\rho_S^i)$. Thus (3) $\Rightarrow$ (1).
		
		It is clear that (4) $\Rightarrow$ (3), since
		\begin{equation}
		(\Phi_A \ox \Phi_B)(\rho_{AB}) = \sum_{N_A,N_B} (P_{N_A} \ox P_{N_B}) \rho_{AB} (P_{N_A} \ox P_{N_B}),
		\end{equation}
		so that if each term in the RHS is separable, then the LHS also is.
		
		Finally, we show that (1) $\Rightarrow$ (4). We have
		\begin{align}
		(P_{N_A} \ox P_{N_B})& \rho_{AB} (P_{N_A} \ox P_{N_B}) \nonumber\\
		&= \sum_i p_i \left( P_{N_A} \rho_A^i P_{N_A} \right) \ox \left( P_{N_B} \rho_B^i P_{N_B} \right),
		\end{align}
		which is separable.
	\end{proof}
\end{lem}
A state can fail to be invariant-separable in two different (but not mutually exclusive) ways: it may break the local SSR, or it may be entangled. The measures of SSR-entanglement defined here capture the amount of entanglement accessible from a single copy of the state under the local SSR. However, there are states which have $E_\text{SSR}=0$ yet are not invariant-separable -- for example, product states which break the local SSR.

\begin{lem}
	The distance-based measure of SSR-entanglement can be calculated by a restricted optimisation over SSR-separable states:
	\begin{equation} \label{eqn:e_ssr_distance_restricted}
	E^D_\text{SSR}(\rho) = \min_{\sigma\in \text{ inv.-sep.}} D(\Phi_A \ox \Phi_B[\rho_{AB}], \sigma_{AB}).
	\end{equation}
	Equivalently, the closest separable state to $(\Phi_A \ox \Phi_B)(\rho_{AB})$ is invariant-separable.
	\begin{proof}
		Let $E'^D_\text{SSR}$ be the quantity defined by the right-hand side of (\ref{eqn:e_ssr_distance_restricted}). We prove an inequality in both directions. Since invariant-separable states form a subset of separable states, it is clear that $E'^D_\text{SSR} \geq E^D_\text{SSR}$. Conversely,
		\begin{align}
		E^D_\text{SSR}(\rho_{AB}) & = \min_{\sigma\in \text{ sep.}} D(\Phi_A \ox \Phi_B[\rho_{AB}] , \sigma_{AB} ) \nonumber \\
		& \geq \min_{\sigma\in \text{ sep.}} D( \Phi_A \ox \Phi_B[\rho_{AB}] , \Phi_A \ox \Phi_B [\sigma_{AB}]) \nonumber \\
		& \geq \min_{\tau\in \text{ inv.-sep.}} D( \Phi_A \ox \Phi_B [\rho_{AB}] , \tau_{AB}) \nonumber \\
		& = E'^D_\text{SSR}(\rho_{AB}),
		\end{align}
		where we have used the monotonicity of $D$ under $\Phi_A \ox \Phi_B$ and the fact that $\Phi_A \ox \Phi_B(\sigma_{AB})$ is invariant-separable.
	\end{proof}
\end{lem}

A useful consequence of Theorem~\ref{thm:block_measures} is that the relative entropy measure of SSR-entanglement can be written as
\begin{align}
E^\text{RE}_\text{SSR}&(\rho_{AB})\nonumber\\
 & = \sum_{N_A,N_B} p_{N_A,N_B} E^\text{RE}_\text{SSR}\left( \frac{(P_{N_A} \ox P_{N_B}) \rho_{AB} (P_{N_A} \ox P_{N_B}) }{p_{N_A,N_B}} \right) \nonumber \\
& = \sum_{N_A,N_B} p_{N_A,N_B} E^\text{RE}\left( \frac{(P_{N_A} \ox P_{N_B}) \rho_{AB} (P_{N_A} \ox P_{N_B}) }{p_{N_A,N_B}} \right),
\end{align}
where $p_{N_A,N_B} = \Tr[(P_{N_A} \ox P_{N_B}) \rho_{AB}]$. This measure is seen to provide an extension of the pure-state measure defined by Wiseman and Vaccaro \cite{wiseman2003entanglement}.

%%%%%%%%%%%%%%%%%%%%%%%%%%%%%%%%%%%%%%%%%%%%%%%%%%%%%%%%%%%%%%%%%%%%%%%%%%%%%%%%%%%%%%%%%%%%%%%%%%%%%%%%%%%%%%%%%%%%%%%%%%
\section{ACTIVATION PROTOCOL} \label{app:activating}
The following Lemma shows that a unitary activation operation can be expressed in a simplified form.
\begin{lem} \label{lem:unitary_activation}
	Let an activation operation $\mc{E}_{C\to AB} \in \mc{O}$ map its input $m$ modes on $C$ directly onto $A$, attach the same number $m$ of vacuum modes in $B$ and interact the two sets by a passive linear unitary $U$:
	\begin{equation}
	\sigma_{AB} = \mc{E}_{C\to AB}(\rho_A) = U(\rho_A \ox {\proj{0}}_B ) U^\dagger.
	\end{equation}
	Up to local free unitaries, $\sigma_{AB}$ is equivalent to the state obtained by replacing $U$ with $D V_A$, where $V_A$ is a free unitary on the $A$ modes and $D$ is a set of beam splitters acting in parallel, with the action
	\begin{equation} \label{eqn:bs_action}
	D^\dagger a_i D = r_i a_i + t_i b_i, \quad r_i =\sqrt{1-t_i^2} \in [0,1],\; i=1,\dots,m.
	\end{equation}
	\begin{proof}
		Lemma 2 of \cite{Yadin2018Operational} shows that $U$ can be decomposed as $W_A W_B D V_A V_B$, where $V_{A,B},\, W_{A,B}$ are free unitaries acting locally on their respective subsystems. Up to final local unitaries, we can replace this by $D V_A V_B$; moreover, $V_B$ can be removed since it leaves the initial vacuum state ${\ket{0}}_B$ unchanged.
	\end{proof}
\end{lem}
It is worth noting that the number of vacuum modes introduced can always be assumed to be no greater than $m$ -- again, as a consequence of Lemma 2 in \cite{Yadin2018Operational}.

The faithfulness of the activation is proven below for almost all such unitaries (apart from those with vanishing beam-splitter parameters).

\begin{manualtheorem}{\ref{thm:activation_faithful}}[{\bf main text}]\label{thm:activation_faithful_X}
	There exists an activation operation $\mc{E}_{C \to AB} \in \mc{O}$ creating an SSR-entangled state $\sigma_{AB}$ from $\rho_C$ if and only if $\rho_C \not \in \mc{S}$.
	
	Moreover, $\mc{E}$ can be taken to be any of the unitary operations described in Lemma~\ref{lem:unitary_activation}, as long as all of the parameters $r_i,t_i$ are non-vanishing.
	\begin{proof}
		
		We first prove that any particle-separable initial state results in no SSR-entanglement. This follows from a more general observation: any bipartite particle-separable state $\rho_{AB}$ also SSR-separable. (This was stated in the two-particle case in Ref.~\cite{wiseman2003entanglement}.) As in the proof of Theorem~\ref{thm:destructive_measurement}, a particle-separable bipartite state ${\ket{\psi}}_{AB}$ can be regarded as an effective two-mode state -- taking $a$ and $b$ as linear combinations of the modes in $A$ and $B$ respectively, we have
		\begin{align}
		{\ket{\psi}}_{AB} & = \left(a^\dagger + b^\dagger \right)^N {\ket{0}}_A {\ket{0}}_B \nonumber \\
		& = \sum_{N_A} r_{N_A} {\ket{N_A}}_A {\ket{N-N_A}}_B,
		\end{align}
		where the $r_{N_A}$ are unimportant coefficients. It is immediate from this expression that $P_{N_A} \ox P_{N-N_A} {\ket{\psi}}_{AB}$ is separable for all $N_A$. Since every particle-separable state is a convex combination of pure particle-separable states, the result follows for all mixed free states. So if $\rho_C$ is a particle-separable state, then for any $\mc{E}_{C \to AB} \in \mc{O}$, $\mc{E}_{C \to AB} (\rho_C)$ is also particle-separable, and hence SSR-separable in the $A/B$ partition.

		Conversely, we prove that any unitary operation as in Lemma~\ref{lem:unitary_activation} with $r_i,t_i \neq 0 \; \forall i$ is sufficient to activate SSR-entanglement from PE. The simplest case -- with a pure state and a ``non-polarising beam-splitter", $r_i = r \, \forall i$ -- was proven in Ref.~\cite{Killoran2014Extracting}. Let us first argue that this extends to mixed states.
		
		Suppose that the output state $\sigma_{AB}$ is SSR-separable, so that each $(P_{N_A} \ox P_{N_B}) \sigma_{AB} (P_{N_A} \ox P_{N_B})$ is separable. As shown in Ref.~\cite{Killoran2014Extracting}, the entanglement structure of $(P_{N_A} \ox P_{N_B}) \sigma_{AB} (P_{N_A} \ox P_{N_B})$ is equivalent to $\rho^{\fq (N_A+N_B)}_{N_A:N_B}$, in which the first-quantised form of the input state is partitioned into $N_A$ versus $N_B$ particles. Hence $\rho^{\fq (N)}$ (with $N=N_A+N_B$) is bi-separable with respect to this partition, i.e.,
		\begin{equation}
		\rho^{\fq (N)} = \sum_i \lambda_i \proj{\phi_i}_{N_A} \ox \proj{\chi_i}_{N_B},
		\end{equation}
		where $\ket{\phi_i} \in \mc{H}_1^{\ox N_A},\, \ket{\chi_i} \in \mc{H}_1^{\ox N_B},\, \lambda_i \geq 0$. Since $\rho^{\fq (N)}$ has support in the symmetric subspace $\mc{H}_N$, we must have ${\ket{\phi_i}}_{N_A} {\ket{\chi_i}}_{N_B} \in \mc{H}_N \, \forall i$. But any bi-separable symmetric pure state must also be fully separable. Therefore ${\ket{\phi_i}}_{N_A} {\ket{\chi_i}}_{N_B} = {\ket{\psi_i}}^{\ox N}$, so $\rho^{\fq (N)}$ is particle-separable.
		
		Finally, we extend to the case of general $r_i$. Via a straightforward generalisation of the argument from Ref.~\cite{Killoran2014Extracting}, we find the output of the activation taking a Fock state $\ket{\bg{n}}$ as input -- the details are in Appendix \ref{app:fock_activation}. Denote by ${\ket{\xi}}_{AB}$ the output of activating $\ket{\bg{n}}$ with beam-splitter parameters $r_i = 1/\sqrt{2} \, \forall i$, and similarly denote by ${\ket{\eta}}_{AB}$ the output obtained with some arbitrary set of $r_i$. From (\ref{eqn:fock_activation}) with two parties and $\alpha_{Ai} = r_i,\, \alpha_{Bi} = t_i$, we have
		\begin{align}
		(P_{N_A} \ox P_{N_B}) &{\ket{\eta}}_{AB}\nonumber\\
		 =& \binom{N}{N_A}^{1/2} \binom{N}{\bg{n}}^{-1/2} \sum_{\substack{\bg{n_A} \\ \sum_i n_{Ai} = N_A \\ n_{Bi}=n_i - n_{Ai} }} \binom{N_A}{\bg{n_A}}^{1/2} \binom{N_B}{\bg{n_B}}^{1/2}\nonumber\\
		 & \left[ \prod_i r_i^{n_{Ai}} t_i^{n_{Bi}} \right] {\ket{\bg{n_A}}}_A {\ket{\bg{n_B}}}_B.
		\end{align}
		It is clear from this expression that $\ket{\eta}$ can be obtained from $\ket{\xi}$ by application of the local operators $L_A \ox L_B$, where
		\begin{align}
		L_A &= \sum_{\bg{n_A}} \left[\prod_i (\sqrt{2} r_i)^{n_{Ai}} \right] \proj{\bg{n_A}},\nonumber\\
		 L_B &= \sum_{\bg{n_B}} \left[\prod_i (\sqrt{2} t_i)^{n_{Bi}} \right] \proj{\bg{n_B}}.
		\end{align}
		Since these operators are independent of the choice of initial Fock state, the same relationship holds for any input state -- that is, the output from an arbitrary set of beam-splitters can be obtained by applying $L_A \ox L_B$ to the output from a set of balanced beam-splitters. As long as $r_i,t_i \neq 0 \, \forall i$, these operators are invertible. The application of invertible local operators to a bipartite state does not change its Schmidt number \cite{Sperling2011Schmidt}. This proves that the faithfulness of activation from a set of arbitrary non-trivial beam-splitters is equivalent to activation from balanced beam-splitters.
	\end{proof}
\end{manualtheorem}

\begin{manualtheorem}{\ref{thm:activation_inequality}}[{\bf main text}]\label{thm:activation_inequality_X}
	For any activation $\mc{E}_{C\to AB} \in \mc{O}, \, E^D_\text{SSR}( \mc{E}_{C\to AB}[\rho_C]) \leq M^D_\text{PE}( \rho_C)$.
	\begin{proof}
		Let $\tau$ be the closest particle-separable state to $\rho$ according to the measure $D$, then
		\begin{align}
		M_{\text{PE}}^D(\rho) & = D(\rho, \tau) \\
		& \geq D( \mc{E}_{C\to AB}(\rho_C), \mc{E}_{C\to AB}(\tau_C)) \\
		& = D(\sigma_{AB}, \mc{E}_{C\to AB}(\tau_C)) \\
		& \geq D\left( \Phi_A \ox \Phi_B (\sigma_{AB}), \Phi_A \ox \Phi_B \circ \mc{E}_{C\to AB}(\tau_C) \right) \\
		& \geq E^D_{SSR}(\sigma_{AB}).
		\end{align}
		The first two inequalities use the contractivity of $D$ under channels. The final inequality uses the fact that $\tau$ is free, so that $\Phi_A \ox \Phi_B \circ \mc{E}_{C\to AB}(\tau_C)$ is separable, but not in general the closest separable state to $\sigma_{AB}$.
	\end{proof}
\end{manualtheorem}

\begin{manualtheorem}{\ref{thm:measure_from_e_ssr_main}}[{\bf main text}]\label{thm:measure_from_e_ssr}
	For any (convex) entanglement measure $E$, the quantity defined as
	\begin{equation}\label{eqn:sup_measure}
	M^E_\text{PE}(\rho) := \sup_{\mc{E}_{C \to AB} \in \mc{O}} E_\text{SSR}\left( \mc{E}_{C \to AB}[\rho_C] \right)
	\end{equation}
	where the supremum is over all deterministic particle-separable operations, is a (convex) measure of PE.
	\begin{proof}
		The faithfulness of the measure is the content of Theorem \ref{thm:activation_faithful}. Deterministic monotonicity follows immediately from the definition and the fact that the set of operations $\mc{O}$ is closed under composition. Non-deterministic (strong) monotonicity states that $M^E_{\text{PE}}\left(\rho\right)$ does not increase on average,
		\begin{align} \label{eqn:ensemble_monotone}
		\sum_i p_i M^E_{\text{PE}}\left(\sigma_i\right)\leq M^E_{\text{PE}}\left(\rho\right)
		\end{align}
		where $\Lambda_i\left(\rho\right)=p_i\sigma_i$ and $\{\Lambda_i\}_i \in \mc{O}$. From the definition \eqref{eqn:sup_measure}, we have, for every activating channel $\mc{E}_{C \to AB} \in \mc{O}$,
		\begin{equation} \label{eqn:non_max_measure}
		M^E_{\text{PE}}(\rho) \geq E_{SSR} \left( \mc{E}_{C \to AB} [\rho_C] \right).
		\end{equation}
		We now continue to prove strong monotonicity by contradiction, showing that a violation of strong monotonicity \eqref{eqn:ensemble_monotone}, implies a violation of \eqref{eqn:non_max_measure}. If strong monotonicity \eqref{eqn:ensemble_monotone} is violated, then there must exist a set of operations $\mc{E}_{i,C\to AB} \in \mc{O}$ such that the following is true:
		\begin{align}\label{eqn:monotonicity_violation}
		M^E_{\text{PE}}(\rho) < \sum_i p_i E_{SSR}\left( \mc{E}_{i,C \to AB}[\sigma_{i,C} ]\right).
		\end{align}
		We now invoke a general property of entanglement measures (and SSR-entanglement measures), namely monotonicity under the partial trace over a subsystem. We split $B$ into two subsystems $B_1, B_2$, in which $B_2$ contains a classical flag. Then, for any ensemble of state $\rho_{i,AB_1}$ with probabilities $p_i$,
		\begin{equation}
		E_{SSR}\left(\sum_i p_i \rho_{i,AB_1}\ox\proj{i}_{B_2}\right)\geq \sum_i p_i E_{SSR}\left(\rho_{i,AB_1}\right).
		\end{equation}
		Applying this to \eqref{eqn:monotonicity_violation}, we obtain
		\begin{align}
		M^E_{\text{PE}}(\rho) & < E_{SSR}\left(\sum_i p_i\mc{E}_{i,C \to AB_1}\left[\sigma_{i,C}\right] \ox\proj{i}_{B_2}\right) \nonumber \\
		& <  E_{SSR}\left(\sum_i \mc{E}_{i,C \to AB_1}\left[\Lambda_{i,C}(\rho_C) \right] \ox\proj{i}_{B_2}\right).
		\end{align}
		
		Note that the operations appearing on the right-hand side above can be combined into a single operation $\mc{F}_{C \to A B_1 B_2} \in \mc{O}$, which is performed by first applying $\{ \Lambda_i \}_i$, storing the outcome $i$ in a classical flag, and then conditionally applying $\mc{E}_i$. Thus,
		\begin{align}
		M^E_{\text{PE}}(\rho) < E_{SSR}\left(\mc{F}_{C\to A B_1 B_2} [\rho_C] \right).
		\end{align}
		The above is a direct contradiction of \eqref{eqn:non_max_measure}, thus establishing that $M_{\text{PE}}$ is a strong monotone for any entanglement monotone $E_{SSR}$.
		
		We now continue by showing convexity:
		\begin{align}
		M^E_{\text{PE}}\left(\sum_i p_i\rho_i\right)\leq\sum_i p_i M_{\text{PE}}\left(\rho_i\right).
		\end{align}
		From the definition of $M_{\text{PE}}$, we have
		\begin{align}
		M^E_{\text{PE}}\left(\sum_i p_i\rho_i\right) & \leq \sup_{\mc{E}_{C\to AB}\in\mc{O}} \sum_i p_i E_{SSR}\left(\mc{E}_{C \to AB}[\rho_{i,C} ]\right)\nonumber\\
		& \leq \sum_i p_i \left\{ \sup_{\mc{E}_{C \to AB}\in\mc{O}} E_{SSR}\left(\mc{E}_{C\to  AB}[\rho_{i,C}] \right) \right\} \nonumber\\
		& = \sum_i p_i M^E_{PE}\left(\rho_i\right).
		\end{align}
		where we have made use of the fact that taking the supremum over each term in the sum individually cannot give less than a single supremum.
	\end{proof}
\end{manualtheorem}
%%%%%%%%%%%%%%%%%%%%%%%%%%%%%%%%%%%%%%%%%%%%%%%%%%%%%%%
\section{LOWER BOUND ON PE MEASURE FROM AN ENTANGLEMENT CRITERION}\label{app:experiment} In order to witness the entanglement present in the system a  criterion of separability from \cite{giovannetti2003characterizing} is used, which is satisfied for all separable states,

\begin{align}\label{cond:sep}
1\leq\frac{4\text{Var}\left( g_z \hat{S}_z^A+\hat{S}_z^B\right)\text{Var}\left( g_y \hat{S}_y^A+\hat{S}_y^B\right)}{\left(\left|g_z g_y\right|\left|\expval{\hat{S}_x^A}\right|+\left|\expval{\hat{S}_x^B}\right|\right)^2},
\end{align}

where $\text{Var}\left(\cdot\right)$ denotes the variance and $g_{(y,z)}$ are real parameters that can be optimised over. The $z$-component of the spin in regions $A,B$  is defined as $\hat{S}^{(A,B)}_{z}:=\frac{1}{2\eta^{(A,B)}_{\text{eff}}}\left(\hat{N}_1^{(A,B)}-\hat{N}_2^{(A,B)}\right)$ where $1,2$ correspond to the internal degree of freedom of the atom and $\eta^{(A,B)}_{\text{eff}}$ accounts for finite spatial resolution in the detection of the BEC. Other spin components, e.g. $\hat{S}^{(A,B)}_x$ and $\hat{S}^{(A,B)}_y$, can be measured by applying appropriate spin rotations before detection. In the following we will show that this condition of separability  \eqref{cond:sep}, can be rewritten as an entanglement witness. \\

Taking the root of equation \eqref{cond:sep} and collecting the terms,
\begin{align}\label{equ:ProdVar}
0\leq& \,\,2\sqrt{\text{Var}\left( g_z \hat{S}_z^A+\hat{S}_z^B\right)\text{Var}\left( g_y \hat{S}_y^A+\hat{S}_y^B\right)}\nonumber\\
&\,\,\,\,\quad\quad\quad\quad-\left(\left|g_z g_y\right|\left|\expval{\hat{S}_x^A}\right|+\left|\expval{\hat{S}_x^B}\right|\right)\nonumber\\
0\leq&\,\,\text{Var}\left( g_z \hat{S}_z^A+\hat{S}_z^B\right)+\text{Var}\left( g_y \hat{S}_y^A+\hat{S}_y^B\right)\nonumber\\ &\,\,\,\,\quad\quad\quad\quad- \left(\left|g_z g_y\right| \expval{\hat{S}_x^A}+\expval{\hat{S}_x^B}\right)\nonumber\\
0\leq&\,\,\text{Var}\left( g_z \hat{S}_z^A+\hat{S}_z^B\right)+\text{Var}\left( g_y \hat{S}_y^A+\hat{S}_y^B\right)\nonumber\\ &\,\,\,\,\quad\quad\quad\quad- \expval{\left|g_z g_y\right| \hat{S}_x^A+\hat{S}_x^B},
\end{align}
where in the second line we have applied the inequality between the geometric and arithmetic mean and removed some of the absolute signs in the third term. We can simplify notation by defining component spin operators  $\hat{S}_z^{+}:=g_z\hat{S}_z^A+\hat{S}_z^B$,   $\hat{S}_y^{+}:=g_y\hat{S}_y^A+\hat{S}_y^B$ and $\hat{S}_x^{+}:=\left|g_z g_y\right|\hat{S}_x^A+\hat{S}_x^B$,
\begin{align} \label{eqn:variance_witness}
\text{Var}\left(\hat{S}_z^{+}\right)+\text{Var}\left(\hat{S}_y^{+}\right)- \expval{\hat{S}_x^{+}}&\geq0.
\end{align}
We now relate this to an entanglement witness observable. For any state $\rho$, let
\begin{align}\label{witness}
W_\rho:=\left(\hat{S}_z^{+}-\expval{\hat{S}_z^{+}}_\rho\right)^2+\left(\hat{S}_y^{+}-\expval{\hat{S}_y^{+}}_\rho\right)^2-\hat{S}_x^{+}.
\end{align}
To check that this is a valid entanglement witness, let $\sigma$ be any separable state. Using $\expval{(X-x_0)^2} = V(X) + (x_0 - \expval{X})^2$, from (\ref{eqn:variance_witness}) we have
\begin{align}
\Tr[\sigma W_\rho] =& \expval{\left(\hat{S}_z^{+}-\expval{\hat{S}_z^{+}}_\rho\right)^2}_\sigma    \nonumber\\
&+\expval{\left(\hat{S}_y^{+}-\expval{\hat{S}_y^{+}}_\rho\right)^2}_\sigma-\expval{\hat{S}_x^{+}}_\sigma\geq0.
\end{align}
Note that when the $\rho$ defining $W_\rho$ is chosen to be the same as the state being measured, the expectation value $\Tr[\rho W_\rho]$ equals the left-hand side of (\ref{eqn:variance_witness}).

Now we have defined an entanglement witness, we can relate such a quantity to a commonly used measure of entanglement defined as the trace distance to the set of separable states,
\begin{align}
M_\text{PE}^{\Tr}(\rho):=\min_{\sigma\in \text{ sep.}}\max_{0\leq P\leq \id}\Tr\left[P(\sigma-\rho)\right],
\end{align}
where $P$ is hermitian. This is by no means the only entanglement measure that can be related to our witness \cite{brandao2005quantifying} but provides a convenient form. As both $P$ and $\sigma$ vary within compact convex sets, and the trace distance is concave for fixed $\sigma$ and convex for fixed $P$, we can make use of the minimax theorem \cite{neumann1928theorie} to obtain
\begin{align}
M_\text{PE}^{\Tr}(\rho)=&\max_{0\leq P\leq \id}\min_{\sigma\in \text{ sep.}}\Tr\left[P(\sigma-\rho)\right].
\end{align}
Now in order to write this measure in terms of the entanglement witness $W_\rho$ we choose a particular $P$:
\begin{equation}
P = W'_\rho + c \id,
\end{equation}
where $c$ is a constant and $W'_\rho = W_\rho/\mc{N}$ is a normalised witness with the factor $\mc{N}$ to be determined later. The constants must be chosen appropriately such that $0 \leq P \leq \id$. This condition is equivalent to
\begin{equation} \label{eqn:normal}
-c \id \leq W'_\rho \leq (1-c) \id,
\end{equation}
which implies that $0<c<1$ since the witness can take values of both signs. Then we have
\begin{align}
M_\text{PE}^{\Tr}(\rho)\geq&\min_{\sigma\in \text{ sep.}}\left[\Tr\left[W_\rho '(\sigma-\rho)\right]+c\Tr\left[\id(\sigma-\rho)\right]\right]\nonumber\\
\geq&-\Tr\left[W_\rho '\rho\right]+\min_{\sigma\in \text{ sep.}}\Tr\left[W_\rho '\sigma\right]\nonumber\\
\geq&-\Tr\left[W_\rho '\rho\right],
\end{align}
where we have used the fact that $\min_{\sigma\in \text{ sep.}}\Tr\left[W'_\rho \sigma\right]\geq0$. \\

We optimise $c$ and $\mc{N}$ to obtain the maximal lower bound on $M_\text{PE}^{\Tr}(\rho)$ subject to normalisation constraints. We start by writing down the range of values taken by the witness,
\begin{align}
W_\rho^{-} \leq \expval{W_\rho} \leq W_\rho^{+},
\end{align}
where $W^{-}_\rho$ and $W^{+}_\rho$ are the minimum and maximum eigenvalues of $W_\rho$. The objective is to make $W_\rho^{-}/\mc{N}$ as negative as possible. Using equation \eqref{eqn:normal}, for given $c$ we want the minimum value of $\mc{N}$ such that $\mc{N}\geq-W^{-}/c$ and $\mc{N}\geq W^{+}/(1-c)$ are both true. We therefore want to choose the normalisation $\mc{N}(c)$ such that
\begin{align}\label{normalisationvalue}
\mc{N}(c)=\max\left\{\frac{-W^{-}}{c},\frac{W^{+}}{1-c}\right\}.
\end{align}
We can see that the minimum value of $\mc{N}(c)$ occurs (for a certain constant $c^*$), when these two terms are equal. We have
\begin{align}
c^{*}=\frac{W^{-}_{\rho}}{W^{-}_\rho -W^{+}_\rho}
\end{align}
and substituting this back into equation \eqref{normalisationvalue} gives us the normalisation constant,
\begin{align}\label{norm}
\mc{N}(c^{*})=W^{+}_{\rho}-W^{-}_\rho.
\end{align}
So the bound on the entanglement measure can therefore be written as,
\begin{align}
M_\text{PE}^{\Tr}(\rho)\geq\frac{-1}{W^{+}_\rho-W^{-}_\rho}\Tr\left[W_\rho \rho\right].
\end{align}
We continue by calculating upper and lower bounds for  $W^{+}_\rho$ and $W^{-}_\rho$ respectively. Starting with $W^{-}_\rho$ we lower bound the product of the variances in the first line of equation \eqref{equ:ProdVar} using the Robertson uncertainty relation,
\begin{align}
\text{Var}\left( g_z \hat{S}_z^A+\hat{S}_z^B\right)&\text{Var}\left( g_y \hat{S}_y^A+\hat{S}_y^B\right)\nonumber\\\geq&\frac{1}{4}\left|\expval{g_z g_y \comm{\hat{S}_z^A}{\hat{S}_y^A}+\comm{\hat{S}_z^B}{\hat{S}_y^B}}\right|^2\nonumber\\
=&\frac{1}{4}\left|\expval{-ig_zg_y\hat{S}_x^A-i\hat{S}_x^B}\right|^2\nonumber\\
=&\frac{1}{4}\expval{g_zg_y\hat{S}_x^A+S_x^B}^2,
\end{align}
where we have used the standard spin commutator relations. This can now be substituted back into the first line of equation \eqref{equ:ProdVar} to lower bound $W_\rho^{-}$ where again we write the second term as a single expectation value,
\begin{align}
W_\rho^{-}\geq\min_\sigma \left[\left|\expval{g_zg_y\hat{S}_x^A+\hat{S}_x^B}_\sigma \right|-\expval{\left|g_z g_y\right| \hat{S}_x^A+\hat{S}_x^B}_\sigma \right]\nonumber\\
\geq0-\max_\sigma \expval{\left|g_z g_y\right| \hat{S}_x^A+\hat{S}_x^B}_\sigma.
\end{align}

The spin operators take their maximal value when all the particles are in internal mode $1$, $\max \hat{S}^{(A,B)}=\frac{1}{2\eta^{(A,B)}_{\text{eff}}}N^{(A,B)}$.
\begin{align}
W_\rho^{-}\geq-\frac{1}{2}\left(\frac{|g_z g_y|N^A}{\eta^{A}_{\text{eff}}}+\frac{N^B}{\eta^{B}_{\text{eff}}}\right),
\end{align}
providing us with a lower bound on $W_\rho^{-}$. We now move onto upper bounding $W_\rho^{+}$. We can start by upper bounding the variance terms in the last line of equation \eqref{equ:ProdVar}. This can be achieved by utilizing Popoviciu's inequality \cite{popoviciu1935equations},
\begin{align}
\text{Var}\left( g_z \hat{S}_z^A+\hat{S}_z^B\right)\leq&\,\frac{1}{4}\left(\lambda_{\max}\left( g_z \hat{S}_z^A+\hat{S}_z^B\right)-\lambda_{\min}\left[ g_z \hat{S}_z^A+\hat{S}_z^B\right]\right)^2\nonumber\\
=&\,\lambda_{\max}\left[g_z \hat{S}_z^A+\hat{S}_z^B\right]^2\nonumber\\
=&\,\left(\abs{g_z}\lambda_{\max}\left[\hat{S}_z^A\right]+\lambda_{\max}\left[\hat{S}_z^B\right]\right)^2\nonumber\\
=&\,\frac{1}{4}\left(\frac{\abs{g_z}N^A}{\eta^{A}_{\text{eff}}}+\frac{N^B}{\eta^{B}_{\text{eff}}}\right)^2
\end{align}
where $\lambda_{\max}[A],\lambda_{\min}[A]$ are the maximum and minimum eigenvalues of the operator $A$, respectively, and in last line we have again used the fact that the value is maximised when all the particles are in the same internal mode. Substituting the above into the last line of equation \eqref{equ:ProdVar} and maximising over each term individually results in,
\begin{align}
W_\rho^{+}\leq&\,\frac{1}{4}\left(\frac{\abs{g_z}N^A}{\eta^{A}_{\text{eff}}}+\frac{N^B}{\eta^{B}_{\text{eff}}}\right)^2+\frac{1}{4}\left(\frac{\abs{g_y}N^A}{\eta^{A}_{\text{eff}}}+\frac{N^B}{\eta^{B}_{\text{eff}}}\right)^2\nonumber\\
&-\min_\sigma \expval{\left|g_z g_y\right| \hat{S}_x^A+\hat{S}_x^B}_\sigma \nonumber\\
\leq&\,\frac{1}{4}\left(\frac{\abs{g_z}N^A}{\eta^{A}_{\text{eff}}}+\frac{N^B}{\eta^{B}_{\text{eff}}}\right)^2+\frac{1}{4}\left(\frac{\abs{g_y}N^A}{\eta^{A}_{\text{eff}}}+\frac{N^B}{\eta^{B}_{\text{eff}}}\right)^2\nonumber\\
&+\frac{1}{2}\left(\frac{|g_z g_y|N^A}{\eta^{A}_{\text{eff}}}+\frac{N^B}{\eta^{B}_{\text{eff}}}\right).
\end{align}
Now we have bounded both the maximum and minimum values the witness can take, we can bound the normalisation $\mc{N}$ from equation \eqref{norm} and therefore bound the entanglement measure with a normalised witness,
\begin{align}\label{eqn:entmlowerbound}
M_\text{PE}^{\Tr}(\rho)\geq&\,-\Bigg{[}\frac{1}{4}\left(\frac{\abs{g_z}N^A}{\eta^{A}_{\text{eff}}}+\frac{N^B}{\eta^{B}_{\text{eff}}}\right)^2+\frac{1}{4}\left(\frac{\abs{g_y}N^A}{\eta^{A}_{\text{eff}}}+\frac{N^B}{\eta^{B}_{\text{eff}}}\right)^2\nonumber\\
&\,+\left(\frac{|g_z g_y|N^A}{\eta^{A}_{\text{eff}}}+\frac{N^B}{\eta^{B}_{\text{eff}}}\right)\Bigg{]}^{-1}\Tr\left[W_\rho \rho\right].
\end{align}
%%%%%%%%%%%%%%%%%%%%%%%%%%%%%%%%%%%%%%%%%%%%%%%%%%%%%%%
\section{NON-CLASSICALITY}\label{appen:non-class}
\begin{thm}
	Every number-diagonal (ND) classical state is particle-separable.
	\begin{proof}
		If $\rho$ is classical and ND, then
		\begin{equation}
		\rho = \int \dd^{2n} \bg{\alpha}\, P(\bg{\alpha}) \Phi(\proj{\bg{\alpha}}),
		\end{equation}
		with $P(\bg{\alpha}) \geq 0$. Hence it is sufficient to prove the claim for all $\Phi(\proj{\bg{\alpha}})$. For any multi-mode coherent state $\ket{\bg{\alpha}}$, there exists a passive linear unitary $U$ that brings all the particles into a single mode: $U\ket{\bg{\alpha}} = \ket{\bar{\alpha}}{\ket{0}}^{\ox(n-1)}$, where $\abs{\bar{\alpha}}^2 = \sum_{i=1}^n \abs{\alpha_i}^2$. Since this unitary is number-conserving, it commutes with $\Phi$, so
		\begin{align}
		U \Phi(\proj{\bg{\alpha}}) U^\dagger & = \Phi \left( U \proj{\bg{\alpha}} U^\dagger \right) \\
		& = \Phi \left( \proj{\bar{\alpha}} \ox {\proj{0}}^{\ox(n-1)} \right) \\
		& = \Phi(\proj{\bar{\alpha}}) \ox {\proj{0}}^{\ox(n-1)} \\
		& = \sum_{k=0}^{\infty} \frac{ e^{-\abs{\bar{\alpha}}^2} \abs{\bar{\alpha}}^{2k} }{k!} \proj{k} \ox {\proj{0}}^{\ox (n-1)},
		\end{align}
		which is particle-separable.
	\end{proof}
\end{thm}

\begin{manualtheorem}{\ref{thm:two_copy_bounded}}[{\bf main text}]\label{thm:two_copy_bounded_X}
	Two copies $\rho^{\ox 2}$ of a number-bounded state $\rho$ are particle-separable if and only if $\rho$ is the vacuum.
	\begin{proof}
		
		Let both $\rho$ and $\rho^{\ox 2}$ be free with bounded particle number, and we decompose $\rho = \sum_{N=0}^{N_0} p_N \rho^{(N)}$. Then
		\begin{equation}
		\rho^{\ox 2} = \sum_{N,N'=0}^{N_0} p_N p_{N'} \rho^{(N)} \ox \rho^{(N')}.
		\end{equation}
		The maximal number component of this state is $p_{N_0}^2 \rho^{(N_0)} \ox \rho^{(N_0)}$, where $p_{N_0} \neq 0$ by assumption. This component must be particle-separable, thus must be obtainable by mixtures of the form $\sum_i p_i U_i {\ket{2N_0, 0, 0, \dots}}{\bra{2N_0, 0, 0, \dots}} U_i^\dagger$, where the $U_i$ are passive linear. Now this state has exactly $N_0$ particles on each of the two parties, and so the same must be true for every term in the sum. In other words, for each $i$, $U_i \ket{2N_0,0} = \left( V_i \ket{N_0}\right) \left(W_i \ket{N_0}\right)$ with pair of additional passive linear unitaries $V_i,W_i$ acting on each subsystem. It is easily seen that this is impossible unless $N_0 = 0$.
	\end{proof}
\end{manualtheorem}

\begin{manualtheorem}{\ref{thm:two_copy_pure}}[{\bf main text}]\label{thm:two_copy_pure_X}
	Two copies $\Phi(\proj{\psi})^{\ox 2}$ of a pseudo-pure state are particle-separable if and only if $\ket{\psi}$ is classical.
	\begin{proof}
		We first show that the activation of an arbitrary pure state $\ket{\psi}$ into SSR-entanglement is exactly the same as for the pseudo-pure state $\Phi(\proj{\psi})$. Let $\Phi_{AB}$ be the joint dephasing operator with respect to the \emph{total} number over two parties $A,B$. This operation is already implemented by dephasing with respect to local number, so that $(\Phi_A \ox \Phi_B) = (\Phi_A \ox \Phi_B) \circ  \Phi_{AB}$. We use this to connect the SSR-entanglement activated by a unitary $\mc{U} \in \mc{O}$ from $\proj{\psi}$ to that activated from $\Phi(\proj{\psi})$:
		\begin{align}
		(\Phi_A \ox& \Phi_B )\circ \mc{U}\big( {\proj{\psi}}_A \ox {\proj{0}_B} \big) \nonumber\\
		& =
		(\Phi_A \ox \Phi_B ) \circ \Phi_{AB} \circ \mc{U} \big( {\proj{\psi}}_A \ox {\proj{0}}_B \big) \\
		& = (\Phi_A \ox \Phi_B) \circ \mc{U} \circ \Phi_{AB} \big( {\proj{\psi}}_A \ox {\proj{0}}_B \big) \\
		& = (\Phi_A \ox \Phi_B) \circ \mc{U} \big( \Phi_A[{\proj{\psi}}_A] \ox {\proj{0}}_B \big),
		\end{align}
		where we have used the fact that $\mc{U}$ is number-conserving, so $[\mc{U},\Phi_{AB}]=0$, and the last line holds because $B$ contains no particles.
		
		Now let $\ket{\psi}$ be activated by $\mc{U}$ consisting of a set of non-trivial beam-splitters into ${\ket{\phi}}_{AB}$. Then we can write ${\ket{\phi}}_{AB} = \sum_{k,l} {\ket{\phi_{k,l}}}_{AB} := \sum_{k,l} P_{k,A} P_{l,B} {\ket{\phi}}_{AB}$. If two copies of $\ket{\psi}$ are activated in the same way in parallel, then the output state is ${\ket{\phi}}^{\ox 2} = {\ket{\phi}}_{A_1B_1} {\ket{\phi}}_{A_2B_2}$. Given that $\Phi(\proj{\psi})^{\ox 2}$ is particle-separable, Theorem \ref{thm:activation_faithful} says that the projection of the activated state onto local particle number must be unentangled -- so there exist (unnormalised) ${\ket{a^{n,m}}}_{A_1A_2} ,\, {\ket{b^{n,m}}}_{B_1B_2}$ such that, for each $n,m$,
		\begin{equation}
		P_{n,A} P_{m,B} {\ket{\phi}}_{A_1B_1}{\ket{\phi}}_{A_2B_2} = {\ket{a^{n,m}}}_{A_1A_2} {\ket{b^{n,m}}}_{B_1B_2}.
		\end{equation}
		Applying the projector $P_{k,A_1} P_{l,B_1}$ onto local numbers in the first copy, we find
		\begin{align}
		\ket{\phi_{k,l}}_{A_1B_1}&{\ket{\phi_{n-k,m-l}}}_{A_2B_2} \nonumber\\
		&= \left( P_{k,A_1} {\ket{a^{n,m}}}_{A_1A_2} \right) \left( P_{l,B_1} {\ket{b^{n,m}}}_{B_1B_2} \right).
		\end{align}
		Both sides of the above equation must be separable with respect to both the $A_1 A_2 / B_1 B_2$ and $A_1 B_1 / A_2 B_2$ partitions. Therefore there must exist (unnormalised) states ${\ket{a^{n,m}_k}}_{A_1},\, {\ket{b^{n,m}_{l}}}_{B_1}$ such that
		\begin{equation}
		{\ket{\phi_{k,l}}}_{A_1B_1} = {\ket{a^{n,m}_k}}_{A_1} {\ket{b^{n,m}_l}}_{B_1}.
		\end{equation}
		The left-hand side of the above is independent of $n$ and $m$, so the same must be true of the states on the right -- removing these labels, we obtain
		\begin{equation}
		{\ket{\phi_{k,l}}}_{A_1B_1} = {\ket{a_k}}_{A_1} {\ket{b_l}}_{B_1}.
		\end{equation}
		Summing over $k$ and $l$, we see that ${\ket{\phi_{k,l}}}_{A_1B_1} = (\sum_k {\ket{a_k}}_{A_1}) (\sum_l {\ket{b_l}}_{B_1})$ is separable. From the result in quantum optics saying that all non-classical states are activated into entangled states, it follows that $\ket{\psi}$ must be classical.
	\end{proof}
\end{manualtheorem}
{

In the following, the vacuum state of any number of modes will be denoted $\ket{0}$. The primitive system $S$ under consideration has $d$ modes, and we denote $k$ copies of $S$ by $S^k$.

The proof of Theorem~\ref{thm:many_copy_bound} relies on the following result, which is of the ``de Finetti" type \cite{caves2002unknown}.

\begin{thm}\label{thm:definetti}
	Let $\rho_{[m]}$ be an exchangeable (i.e., permutation-symmetric) state of $N$ particles on $m$ modes that is also particle-separable. Denote by $\rho_{[l]}$ the reduced state of any subset of $l \leq m$ modes. Then there exists a classical $l$-mode state $\sigma_{[l]}$ such that
	\begin{equation}
		D_{\Tr}(\rho_{[l]}, \sigma_{[l]}) \leq \frac{l}{m},
	\end{equation}
	\begin{proof}
		Since $\rho_{[m]}$ is particle separable, there is a probability distribution  $q_\lambda$ and a set of single-particle creation operators $c_\lambda^\dagger$ such that
		\begin{equation}
			\rho_{[m]} = \sum_\lambda \frac{q_\lambda}{N!} (c_\lambda^\dagger)^N \proj{0} c_\lambda^N.
		\end{equation}
		We decompose $c_\lambda^\dagger = \alpha_\lambda a_\lambda^\dagger + \alpha'_\lambda {a'_\lambda}^\dagger$, where $\abs{\alpha_\lambda}^2 + \abs{\alpha'_\lambda}^2 = 1$, $a_\lambda$ acts on modes $1,\dots,l$, and $a'_\lambda$ acts on modes $l+1,\dots,m$.
		Using the binomial expansion for $(c_\lambda^\dagger)^N$ and tracing out modes $l+1,\dots,m$, we have
		\begin{align}
			\rho_{[l]} & = \Tr_{l+1,\dots,m} \rho_{[m]} \nonumber \\
				& = \sum_{n=0}^N \frac{1}{N!} \binom{N}{n}^2 \abs{\alpha_\lambda}^{2n} \abs{\alpha'_\lambda}^{2(N-n)} (a_\lambda^\dagger)^n \proj{0} a_\lambda^n \nonumber \\
				& = \sum_{n=0}^N \binom{N}{n} \abs{\alpha_\lambda}^{2n} (1- \abs{\alpha_\lambda}^2)^{N-n} \proj{n^{(\lambda)}} \nonumber \\
				& = \sum_{n=0}^N b_\lambda(n) \proj{n^{(\lambda)}},
		\end{align}
		where $b_\lambda$ is the binomial distribution with $N$ trials and $p = \abs{\alpha_\lambda}^2$, and $\ket{n^{(\lambda)}} := \frac{1}{\sqrt{n!}} (a_\lambda^\dagger)^n \ket{0}$.
		
		Now we use a result on the Poisson distribution as a limit case of the binomial distribution. For a binomial $b(n)$ and Poisson $\pi(n)$ with the same mean $\mu$, it is well known that $b \to \pi$ in the limit of large $N$. In fact, a stronger result \cite{Harremoes2004Rate}(Eq.~4) says that
		\begin{equation} \label{eqn:bin_pois_inequality}
			D_{\Tr}(b, \pi) \leq p = \frac{\mu}{N},
		\end{equation}
		where $D_{\Tr}$ here is the classical version of the trace distance.
		
		Let $\pi_\lambda$ be the Poisson distribution with mean $\mu_k = N \abs{\alpha_\lambda}^2$, and define
		\begin{equation}
			\sigma_{[l]} := \sum_\lambda q_\lambda \sum_{n=0}^\infty \pi_\lambda(n) \proj{n^{(\lambda)}}.
		\end{equation}
		Note that $\sigma_{[l]}$ is classical since it can be written in the form
		\begin{align}
			\sigma_{[l]} & = \sum_\lambda q_\lambda \Phi(\proj{\psi_\lambda}), \\
			\ket{\psi_\lambda} & := \sum_{n=0}^\infty \sqrt{\pi_\lambda(n)} \ket{n^{(\lambda)}},
		\end{align}
		where $\ket{\psi_\lambda}$ is a coherent state with mean particle number $\mu_\lambda$. It follows that
		\begin{align}
			D_{\Tr}(\rho_{[l]}, \sigma_{[l]}) & = \frac{1}{2} \left\| \sum_{\lambda,n} q_\lambda [b_\lambda(n) - \pi_\lambda(n)] \proj{n^{(\lambda)}} \right\|_{1} \nonumber \\
				& \leq \frac{1}{2} \sum_{\lambda,n} q_\lambda \left\| [b_\lambda(n) - \pi_\lambda(n)] \proj{n^{(\lambda)}} \right\|_1 \nonumber \\
				& = \sum_\lambda q_\lambda \sum_n \frac{1}{2} \abs{b_\lambda(n) - \pi_\lambda(n)} \nonumber \\
				& = \sum_\lambda q_\lambda D_{\Tr}(b_\lambda, \pi_\lambda) \nonumber \\
				& \leq \sum_\lambda q_\lambda \frac{\mu_\lambda}{N},
		\end{align}
		having used the triangle inequality and finally \eqref{eqn:bin_pois_inequality}. Now $\sum_\lambda q_\lambda \mu_\lambda$ is the mean particle number in $\rho_{[l]}$, which by exchangeability is $N l/m$. Therefore
		\begin{equation}
			D_{\Tr}(\rho_{[l]}, \sigma_{[l]}) \leq \frac{l}{m}.
		\end{equation}
	\end{proof}
\end{thm}

\begin{manualtheorem}{\ref{thm:many_copy_bound}}[{\bf main text}]
    Let $\rho$ have finite mean particle number, $\Tr[\rho \hat{N}] < \infty$, and suppose that $\rho^{\ox k}$ is particle-separable for some $k$. Then the trace-distance non-classicality of $\rho$ is bounded by
    \begin{equation}
        M^{\Tr}_\mathrm{NC}(\rho) \leq \frac{1}{k}.
    \end{equation}
    Consequently, $\rho^{\ox k}$ is particle-separable for all $k$ if and only if $\rho$ is classical.
    \begin{proof}
        Let $\rho$ contain $d$ modes, so that $\rho^{\ox k}$ contains $m = kd$ modes. Projecting onto the subspace of total particle number $N$ results in the (normalised) state $P_{N,S^k} \rho^{\ox k} P_{N,S^k} / p_N$, which fulfils the assumptions of Theorem~\ref{thm:definetti}. Therefore there exists a classical state $\sigma_N$ of $d$ modes such that
		\begin{equation}
			D_{\Tr}\left( \frac{ \Tr_{S_2,\dots,S_k} P_{N,S^k} \rho^{\ox k} P_{N,S^k}}{p_N} , \sigma_N \right) \leq \frac{d}{kd} = \frac{1}{k}.
		\end{equation}
		
		Defining the classical state $\sigma := \sum_N p_N \sigma_N$, we have
		\begin{align}
			D_{\Tr} (\rho, \sigma) & = D_{\Tr} \left( \sum_N  \Tr_{S_2,\dots,S_k} P_{N,S^k} \rho^{\ox k} P_{N,S^k}, \sum_N p_N \sigma_N \right) \nonumber \\
				& \leq \sum_N p_N D_{\Tr} \left( \frac{ \Tr_{S_2,\dots,S_k} P_{N,S^k} \rho^{\ox k} P_{N,S^k}}{p_N}, \sigma_N \right) \nonumber \\
				& \leq \sum_N p_N \frac{1}{k} \nonumber \\
				& = \frac{1}{k},
		\end{align}
		having used convexity of $D_{\Tr}$.
		
		The final statement is an immediate application of this bound in the limit $k \to \infty$, using the fact that the set of classical states is closed in the trace-norm topology \cite{Bach1986Simplex}. Conversely, it is enough to note that the set of classical states is closed under tensor products.
    \end{proof}
\end{manualtheorem}
}

\section{UNITARY ACTIVATION OF FOCK STATES} \label{app:fock_activation}
Here we generalise the main result of Ref.~\cite{Killoran2014Extracting} to multiple modes and to general beam-splitters. We also present the results without much additional effort for arbitrary numbers of parties, although the rest of our work uses only the bipartite case. Let us first find the first-quantised form of an $m$-mode Fock state $\ket{\bg{n}}$, partitioned into sets of $N_A,N_B,\dots,N_Z$ particles, where $\sum_{K=A,B,\dots,Z} N_K = N := \sum_i n_i$. We have
\begin{equation}
{\ket{\bg{n}}}^{\fq} = \binom{N}{\bg{n}}^{-1/2} \sum_\Pi \Pi \bigotimes_{i=0}^{m-1} {\ket{i}}^{\ox n_i},
\end{equation}
where $\binom{N}{\bg{n}}$ is a multinomial coefficient and the sum runs over \emph{distinct} permutations $\Pi$ of $\bigotimes_{i=0}^{m-1} {\ket{i}}^{\ox n_i}$. Dividing initially into $N_A$ versus $N_{\bar{A}} = N - N_A$ particles, it may be verified that
\begin{equation}
{\ket{\bg{n}}}^{\fq} = \binom{N}{\bg{n}}^{-1/2} \sum_{\substack{\{n_{Ai}\}_i \\ \sum_i n_{Ai}=N_A}} \binom{N_A}{\bg{n_A}}^{1/2} \binom{N_{\bar A}}{\bg{n_{\bar A}}}^{1/2} {\ket{\bg{n_A}}}^{\fq}_{N_A} {\ket{\bg{n_{\bar A}}}}^{\fq}_{N_{\bar A}},
\end{equation}
where $n_{\bar{A}i} = n_i - n_{Ai}$. Recursively continuing the subdivision of $\bar A$ in this way, we obtain
\begin{equation} \label{eqn:fock_divided_form}
{\ket{\bg{n}}}^{\fq} = \binom{N}{\bg{n}}^{-1/2} \sum_{\substack{\{ \bg{n_K} \}_K \\ \sum_i n_{Ki} = N_K \forall K \\ \sum_K n_{Ki} = n_i \forall i}} \bigotimes_K \binom{N_K}{\bg{n_K}}^{1/2} {\ket{\bg{n_K}}}^{\fq}_{N_K}.
\end{equation}

Next, we show how a Fock state on $A$ is activated into a multipartite SSR-entangled state by mixing with vacuum modes on $B,\dots,Z$ at a generalised beam splitter. Specifically, we take the beam-splitter $U$ to have the action $a_{Ai}^\dagger \to \sum_K \alpha_{Ki} a_{Ki}^\dagger$ -- a generalisation of Ref.~\cite{Killoran2014Extracting}, in which $\alpha_{Ki}$ was independent of $i$. Then
\begin{align}
{\ket{\phi}}&_{A\dots Z}:= U {\ket{\bg{n}}}_A {\ket{00\dots}}_{B\dots Z} \nonumber \\
=& \prod_i \frac{1}{\sqrt{n_i!}} \left( \sum_K \alpha_{Ki} a_{Ki}^\dagger \right)^{n_i} {\ket{00\dots}}_{A\dots Z} \nonumber \\
 =& \prod_i \frac{1}{\sqrt{n_i!}} \sum_{\substack{\{ n_{Ki} \} \\ \sum_K n_{Ki} = n_i \forall i}} \binom{n_i}{n_{Ai},\dots,n_{Zi}}\nonumber\\
&\quad\quad \prod_K (\alpha_{Ki} a_{Ki}^\dagger)^{n_{Ki}} {\ket{00\dots}}_{A\dots Z} \nonumber \\
=& \sum_{\substack{ \{\bg{n_K}\}_K \\ \sum_K n_{Ki} = n_i \forall i}} \left[ \prod_i \binom{n_i}{n_{Ai},\dots,n_{Zi}}^{1/2} \right]\nonumber\\
& \quad\quad\bigotimes_K  \left[ \prod_i \alpha_{Ki}^{n_{Ki}} \right] {\ket{\bg{n_K}}}_K .
\end{align}
Conditioning on local particle number,
\begin{align} \label{eqn:fock_activation}
(P&_{N_A} \ox \dots \ox P_{N_Z}) {\ket{\phi}}_{A\dots Z} \nonumber\\ =& \sum_{\substack{ \{\bg{n_K}\}_K \\ \sum_i n_{Ki} = N_K \forall K \\ \sum_K n_{Ki} = n_i \forall i}} \left[ \prod_i \binom{n_i}{n_{Ai},\dots,n_{Zi}}^{1/2} \right] \bigotimes_K  \left[ \prod_i \alpha_{Ki}^{n_{Ki}} \right] {\ket{\bg{n_K}}}_K \nonumber \\
 =& \left[ \frac{\prod_i n_i!}{\prod_K N_K!} \right]^{1/2}  \sum_{\substack{ \{\bg{n_K}\}_K \\ \sum_i n_{Ki} = N_K \forall K \\ \sum_K n_{Ki} = n_i \forall i}} \bigotimes_K \binom{N_K}{\bg{n_K}}^{1/2} \left[ \prod_i \alpha_{Ki}^{n_{Ki}} \right] {\ket{\bg{n_K}}}_K \nonumber \\
  = &\binom{N}{N_A,\dots,N_Z}^{1/2} \binom{N}{\bg{n}}^{-1/2} \sum_{\substack{ \{\bg{n_K}\}_K \\ \sum_i n_{Ki} = N_K \forall K \\ \sum_K n_{Ki} = n_i \forall i}} \bigotimes_K \binom{N_K}{\bg{n_K}}^{1/2}\nonumber\\
  &\quad\quad\quad\quad \quad\quad\quad\quad\quad\quad\,\,\left[ \prod_i \alpha_{Ki}^{n_{Ki}} \right] {\ket{\bg{n_K}}}_K,
\end{align}
which is of the same form as (\ref{eqn:fock_divided_form}), up to the coefficients $\binom{N}{N_A,\dots,N_Z}^{1/2} \prod_{K,i} \alpha_{Ki}^{n_{Ki}}$.

%%TC:endignore
%%%%%%%%%%%%%%%%%%%%%%%%%%%%%%%%%%%%%%%%%%%%%%%%%%%%%%%%%%%%%%%%%%%%%%%%%%%%
%\newpage
% Don't count these!
%%TC:ignore
%\detailtexcount{Naturedraftv6}
%%TC:endignore
\end{document}